\documentclass[11pt]{article}
\usepackage{amsmath}
\normalbaselines

\newtheorem{thm}{Theorem}[section]
\newtheorem{lem}{Lemma}[section]

\newtheorem{cor}{Corollary}[section]

\begin{document}
\bibliographystyle{unsrt}

\def\bea*{\begin{eqnarray*}}
\def\eea*{\end{eqnarray*}}
\def\ba{\begin{array}}
\def\ea{\end{array}}
\count1=1
\def\be{\ifnum \count1=0 $$ \else \begin{equation}\fi}
\def\ee{\ifnum\count1=0 $$ \else \end{equation}\fi}
\def\ele(#1){\ifnum\count1=0 \eqno({\bf #1}) $$ \else \label{#1}\end{equation}\fi}
\def\req(#1){\ifnum\count1=0 {\bf #1}\else \ref{#1}\fi}
\def\bea(#1){\ifnum \count1=0   $$ \begin{array}{#1}
\else \begin{equation} \begin{array}{#1} \fi}
\def\eea{\ifnum \count1=0 \end{array} $$
\else  \end{array}\end{equation}\fi}
\def\elea(#1){\ifnum \count1=0 \end{array}\label{#1}\eqno({\bf #1}) $$
\else\end{array}\label{#1}\end{equation}\fi}
\def\cit(#1){
\ifnum\count1=0 {\bf #1} \cite{#1} \else 
\cite{#1}\fi}
\def\bibit(#1){\ifnum\count1=0 \bibitem{#1} [#1    ] \else \bibitem{#1}\fi}
\def\ds{\displaystyle}
\def\hb{\hfill\break}
\def\comment#1{\hb {***** {\em #1} *****}\hb }

\newcommand{\TZ}{\hbox{\bf T}}
\newcommand{\MZ}{\hbox{\bf M}}
\newcommand{\ZZ}{\hbox{\bf Z}}
\newcommand{\NZ}{\hbox{\bf N}}
\newcommand{\RZ}{\hbox{\bf R}}
\newcommand{\CZ}{\,\hbox{\bf C}}
\newcommand{\PZ}{\hbox{\bf P}}
\newcommand{\QZ}{\hbox{\rm eight}}
\newcommand{\HZ}{\hbox{\bf H}}
\newcommand{\EZ}{\hbox{\bf E}}
\newcommand{\GZ}{\,\hbox{\bf G}}

\font\germ=eufm10
\def\goth#1{\hbox{\germ #1}}
\vbox{\vspace{38mm}}

\begin{center}
{\LARGE \bf On the Equivalent Theory of the Generalized $\tau^{(2)}$-model and the Chiral Potts Model with two Alternating Vertical Rapidities } \\[10 mm] 
Shi-shyr Roan \\
{\it Institute of Mathematics \\
Academia Sinica \\  Taipei , Taiwan \\
(email: maroan@gate.sinica.edu.tw ) } \\[25mm]
\end{center}

\begin{abstract}
By the Baxter's $Q_{72}$-operator method, we demonstrate the equivalent theory between the generalized $\tau^{(2)}$-model (other than two special cases  with a pseudovacuum state) and  the $N$-state chiral Potts model with two alternating vertical rapidities, where the degenerate models are included. As a consequence, the theory of the XXZ chain model associated to  cyclic representations (with the parameter $\varsigma$) of $U_{\sf q}(sl_2)$ with ${\sf q}^N=1$ for odd $N$ is identified with either (for $\varsigma^N=1$) the chiral Potts model with two superintegrable vertical rapidities, or (for $\varsigma^N \neq 1$) the degenerate model for the selfdual solution of the star-triangle relation. In all these identifications, the transfer matrices $T, \widehat{T}$ of the chiral Potts model (including the degenerate ones) serve as the $Q_R, Q_L$-operators of the corresponding $\tau^{(2)}$-model, so that the functional relations hold as in the solvable $N$-state chiral Potts model.

\end{abstract}
\par \vspace{5mm} \noindent
{\rm 2006 PACS}:  05.50.+q, 03.65.Fd, 75.10.Pq \par \noindent

\setcounter{section}{0}
\section{Introduction}
\setcounter{equation}{0}
In the study of $N$-state chiral Potts model (CPM) as a descendant of the six-vertex model,  Bazhanov and Stroganov \cite{BazS} found a five-parameter family of Yang-Baxter (YB) solutions for the asymmetric six-vertex $R$-matrix, which defines the generalized $\tau^{(2)}$-model, also known as the Baxter-Bazhanov-Stroganov model \cite{B89, B049, BazS, GIPS}.  
The transfer chiral-Potts matrix arises as the $Q$-operator of the corresponding $\tau^{(2)}$-matrix  \cite{BBP, BazS} by following the construction of Baxter's $Q$-operator for the eight-vertex model in \cite{Bax}. Hereafter in this paper, the CPM  always means the "checkerboard" type model with two vertical (alternating) rapidities as discussed in \cite{BBP}, where the functional-relation method was invented due to the lack of the "difference" property of CPM rapidities in a high-genus curve. By counting the free parameters of CPM, one easily see that the $\tau^{(2)}$-models arisen from CPM form a three-parameter sub-family among all generalized $\tau^{(2)}$-models. 
The aim of this paper is to conduct the $Q$-operator investigation for an arbitrary generalized $\tau^{(2)}$-model along the line of Baxter's $Q_{72}$-operator in the eight-vertex model \cite{B72}. First, we note that a pseudovacuum state exists only for a certain special type of $\tau^{(2)}$-models, which can be studied by the powerful algebraic Bethe ansatz method \cite{Fad, KBI, KS, TakF} as previously shown in \cite{R06F}. Except those $\tau^{(2)}$-models possessing a pseudovacuum state, the main result of this paper can be loosely stated as " the generalized $\tau^{(2)}$-models and CPM with two vertical rapidities are the equivalent theories provided degenerate versions of CPM are included". The CPM transfer matrix will be derived as the $Q$-operator of the corresponding $\tau^{(2)}$-model in the functional-relation framework \cite{BBP, R05b}. Note that the $\tau^{(2)}$-model in this work is the trace of product of $L$-operator (\req(Lab')), which is invariant under gauge and scale transforms (\req(nu)) (\req(lambda)). Using these transformations, one can always reduce the $\tau^{(2)}$-model to one in CPM with the alternating rapidities having the same temperature-like parameter $k'$. Hence the $Q$-operator is the CPM  transfer matrix in \cite{BBP},  however the degenerate forms are necessarily included. 
Furthermore,  Baxter extended the study of CPM transfer matrix and functional relations to some $\tau^{(2)}$-models \cite{B049} more general than those considered in this work. The generalized $\tau^{(2)}$-model of Baxter in \cite{B049} is an "inhomogeneous" model of  alternating rapidities with two $k'$s, not generally equal  even by the gauge and scale transforms.  Then the Boltzmann weights not necessarily satisfy the usual $N$-periodicity conditions, but replaced by a weaker condition (\cite{B049} (27)).

By the observation that a special gauge transformation and the rescaling of spectral variables of the $L$-operator give rise to the equivalent $\tau^{(2)}$-models, a "generic" $\tau^{(2)}$-model can be reduced to a $\tau^{(2)}$-model in CPM. Indeed, one can derive the quantitative description of the "generic"-criterion about parameters in $L$-operator by the algebraic geometry study of these equivalent relations among $\tau^{(2)}$-models. As a consequence of this result, the conjectural boundary fusion relation  \cite{GIPS, R06F} holds for an arbitrary generalized $\tau^{(2)}$-model, hence the method of separation of variables can be applied in the study of $\tau^{(2)}$-models (\cite{GIPS} Theorem 2).
Furthermore, the non-generic $\tau^{(2)}$-models are now the only remaining cases where an appropriate $Q$-operator is to be found in the theory.  In this paper, we employ the Baxter's techniques of producing $Q_{72}$-operator of the root-of-unity eight-vertex model \cite{B72} to construct the $Q_R, Q_L$-, then $Q$-operator for a given $\tau^{(2)}$-model, as in the $Q$-operator study of  the root-of-unity XXZ and eight-vertex model in \cite{B72, B73, DFM, FM01, FM02, FM04, F06, R06Q, R06Q8, R07}, also as the superintegrable CPM in \cite{R075}. Indeed, we show that the method can be successfully applied  to CPM with two arbitrary vertical rapidities  to reproduce the CPM transfer matrices $T, \widehat{T}$, originally appeared in \cite{BBP}, from the $\tau^{(2)}$-matrix as its $Q_R, Q_L$-operators.
The special (i.e. non-generic) $\tau^{(2)}$-models, other than two special cases (see (\req(aba)) in the paper) where the pseudovacuum state exists, are now reduced to the theory of "degenerate chiral Potts models" for $k'=1, 0$. The degenerate chiral Potts models for $k'=1$ are indeed the selfdual solutions of the star-triangle relation in  \cite{AMPTY, AMPT, MPTS, BPA, FatZ}. Consequently,  the theory of XXZ chains associated to  cyclic representations (with the parameter $\varsigma$) of $U_{\sf q}(sl_2)$ for ${\sf q}^N=1$ and odd $N$ can be identified with either the superintegrable CPM with two vertical rapidities (for $\varsigma^N=1$), or  the selfdual Potts model in \cite{BPA, FatZ} (for $\varsigma^N \neq 1$). Among these identifications is the equivalent theory of the spin-$\frac{N-1}{2}$ XXZ chain for ${\sf q}^N=1$ and a homogeneous superintegrable CPM, as  previously shown in \cite{R075}. Furthermore, the $\tau^{(2)}$-matrix of a degenerate chiral Potts model is explicitly given in our approach so that the whole set of functional relations in CPM \cite{BBP} holds also in the degenerate model. This suggests that one should be able to carry out a study of the degenerate chiral Potts model (but not done yet) on various problems, such as the eigenvalue spectrum of the transfer matrix, similar to those in the solvable $N$-state chiral Potts model in \cite{B90, MR}.

This paper is organized as follows. In section \ref{sec:CT}, we briefly review some basic facts in CPM and the generalized $\tau^{(2)}$-model. First we recall known results in CPM in subsection \ref{ssec.CP}, then give a brief discussion of  the generalized $\tau^{(2)}$-model in subsection \ref{ssec.tau2}. Here we state one of main results in this paper, Theorem \ref{thm:CPt}, about the precise criterion of $\tau^{(2)}$-models equivalent to those in CPM with two vertical rapidities. The proof of Theorem \ref{thm:CPt} is based on an algebraic geometry study of rapidity curves for $k' \neq 0, \pm 1$, the detailed argument of which we leave in Appendix where some technical complexity in mathematical derivation seems necessary due to the constraint of the parameter $k'$. In section \ref{ssec.QtCP}, we provide a construction of $Q$-operator of a generalized $\tau^{(2)}$-model using the Baxter's $Q_{72}$-operator method \cite{B72}. We illustrate this construction by reproducing the chiral Potts transfer matrices from the $\tau^{(2)}$-model as its $Q_R$ and $Q_L$-operators. The method will also enable us to derive the degenerate chiral Potts models from the $\tau^{(2)}$-model not covered in Theorem \ref{thm:CPt}. In section \ref{ssec.sdCP}, the selfdual degenerate Potts model with $k'=1$ (in \cite{AMPT, MPTS, FatZ}, \cite{BPA}(10)) are found  through the $Q$-operator theory of certain $\tau^{(2)}$-models, among which are those equivalent to XXZ chains for cyclic representations of $U_{\sf q}(sl_2)$ previously described in \cite{R075} with ${\sf q}^N=1$ and representation parameter $\varsigma^N \neq 1$ for odd $N$. In section \ref{sec:dCP}, we study the $Q$-operator theory of the remaining special $\tau^{(2)}$-models (with conditions  (\req(nz)) (\req(sCn1)), and ${\sf c}^N \neq 1$). The $Q_R$, $Q_L$-operators are constructed through the theory of degenerate chiral Potts models with $k'=1, 0$. However, the commutating relation between $Q_R$ and $Q_L$ required for the construction of commuting $Q$-operators holds only in case $k'=1$, which is studied in subsection \ref{ssec.k'1}. The degenerate chiral Potts models with $k'= 0$ is discussed in subsection \ref{ssec.k'0}.  Since the Boltzmann weights of each case for degenerate chiral Potts model with $k'=1$ give arise to a solution of the star-triangle relation, we observe that the derivation of functional relations in CPM  in \cite{BBP} holds also for these degenerated models. The functional relations of all those models are listed in section \ref{sec:FRdCP}. We close in section \ref{sec.F} with some concluding remarks.

\section{Chiral Potts Model and the Generalized $\tau^{(2)}$-model \label{sec:CT} } 
\setcounter{equation}{0}
This section serves as a brief introduction to the chiral Potts model and the generalized $\tau^{(2)}$-model with a sketchy summary, also used for establishing the notation (for more details, see \cite{AMP, BBP, BazS, R04, R05o} and references therein).  In subsection \ref{ssec.tau2}, we describe a precise $\tau^{(2)}$-matrix criterion of CPM with two vertical rapidities among all generalized $\tau^{(2)}$-models, stated as Theorem \ref{thm:CPt}, the proof of which  we leave in Appendix.

In this paper, $\CZ^N$ denotes the vector space of $N$-cyclic vectors $v = \sum_{n \in \ZZ_N} v_n | n \rangle$ with the basis indexed by  $n \in \ZZ_N (:= \ZZ/N\ZZ)$. We fix the $N$th root of unity $\omega = e^{\frac{2 \pi \sqrt{-1}}{N}}$, and a pair of Weyl $\CZ^N$-operators, $X$ and  $Z$, with the relations $XZ= \omega^{-1}ZX$ and  $X^N=Z^N=1$:
$$
 X |n \rangle = | n +1 \rangle , ~ \ ~ Z |n \rangle = \omega^n |n \rangle ~ ~ \ ~ ~ (n \in \ZZ_N) .
$$

\subsection{ The $N$-state chiral Potts model \label{ssec.CP}}
The rapidities  of the $N$-state CMP are elements of a genus-$(N^3-2N^2+1)$ curve described by the four-vector ratios $[a, b, c, d]$ in the projective 3-space $ \PZ^3$ with the equation
\be
{\goth W}_{k'}  : \left\{ \begin{array}{ll}
ka^N + k'c^N = d^N , &
kb^N + k'd^N = c^N ,  \\
a^N + k'b^N = k d^N , &
k'a^N + b^N =
k c^N , \end{array} \right.
\ele(rapidC)
where $k, k'$ are parameters with $k^2 + k'^2 = 1$, and $k' \neq \pm 1, 0$. Note that the four relations in (\req(rapidC)) are determined by an arbitrary two among them. In this paper, the variables $x, y, \mu , t$ will denote the following component-ratios of $[a, b, c, d] \in \PZ^3$:
$$
x := \frac{a}{d} , \ \ \  y :=  \frac{b}{c} , \ \ \ \mu : =   \frac{d}{c} , \ \ \ \ t : = \frac{a b}{c d } \ (= x y ).
$$
For later use, we define the following $\PZ^3$-automorphisms:
\bea(lll)
R : (x, y, \mu ) \mapsto (y, \omega x, \mu^{-1} ) , &
T : (x, y, \mu ) \mapsto (\omega x, \omega^{-1} y, \omega^{-1}\mu ) , &
U : (x, y, \mu ) \mapsto (\omega x, y, \mu ).
\elea(Aut)
Note that the above automorphisms leave the curve ({\req(rapidC)) unchanged.
Hereafter, we shall use letters $p, q, \ldots$ to denote elements in $\PZ^3$, and write its ratio-coordinates by $x_p, y_p, t_p, \mu_p, a_p, b_p, \ldots $ whenever it will be necessary to specify the element $p$. 
Using the coordinates $(x , y , \mu ) \in \CZ^3$, the curve ${\goth W}_{k'}$ (\req(rapidC)) is defined by the equation:
\be
{\goth W}_{k'}: ~ ~ k x^N  = 1 -  k'\mu^{-N}, \ ~ \  k  y^N  = 1 -  k'\mu^N ,
\ele(xymu)
which is equivalent to 
\be
k = \frac{x^N + y^N}{1 + x^N y^N }, \  \mu^{-N}   = \frac{1 -  k x^N}{k'} \ \bigg(\Longleftrightarrow \ k = \frac{x^N + y^N}{1 + x^N y^N }, \ \ \mu^N   = \frac{1 -  k y^N}{k'}\bigg) .
\ele(xyx)
The condition for $k \neq 0, \pm 1, \infty $ is equivalent to the constraint: either $(x^N + y^N)(1-x^{2N})(1-y^{2N})(1+x^Ny^N) \neq 0$, or  $x^N = - y^N = \pm 1 $. 
The Boltzmann weights of the $N$-state CPM are defined by coordinates of $p, q \in {\goth W}_{k'}$: 
\bea(ll)
\frac{W_{p,q}(n)}{W_{p,q}(0)}  = (\frac{\mu_p}{\mu_q})^n \prod_{j=1}^n
\frac{y_q-\omega^j x_p}{y_p- \omega^j x_q } &= \prod_{j=1}^n
\frac{d_pb_q-a_pc_q\omega^j}{b_pd_q-c_pa_q\omega^j} , \\
\frac{\overline{W}_{p,q}(n)}{\overline{W}_{p,q}(0)}  = ( \mu_p\mu_q)^n \prod_{j=1}^n \frac{\omega x_p - \omega^j x_q }{ y_q- \omega^j y_p } & = \prod_{j=1}^n
\frac{\omega a_pd_q-
d_pa_q\omega^j}{ c_pb_q- b_pc_q \omega^j},
\elea(CPW)
which satisfy the star-triangle relation 
\be
\sum_{n=0}^{N-1} \overline{W}_{qr}(j' - n) W_{pr}(j - n) \overline{W}_{pq}(n - j'')= R_{pqr} W_{pq}(j - j')\overline{W}_{pr}(j' - j'') W_{qr}(j -j'')    
\ele(TArel)
where $R_{pqr}= \frac{f_{pq}f_{qr}}{f_{pr}}$ with $f_{pq} : = \bigg( \frac{{\rm det}_N( \overline{W}_{pq}(i-j))}{\prod_{n=0}^{N-1} W_{pq}(n)}\bigg)^{\frac{1}{N}}$. Note that
the rapidity constraint (\req(xymu)) ensures the Boltzmann weights (\req(CPW)) with the $N$-periodic property for $n$.  
On a lattice of the horizontal size $L$, the combined weights of
intersections with vertical rapidities $p, p'$ between two consecutive rows give rise
to the operator of $\stackrel{L}{\otimes} \CZ^N$ which defines the transfer matrix of the $N$-state CPM:
\be
T_{p,p'} (q)_{\{j \}, \{j'\}} = \prod_{\ell =1}^L W_{p,q}(j_\ell - j'_\ell )
\overline{W}_{p',q}(j_{\ell+1} - j'_\ell),
\ele(Tpq)
for $q \in {\goth W}_{k'}$ and  $j_\ell, j'_\ell \in \ZZ_N$. Here the periodic condition is imposed by defining $L+1=1$. Similarly, we define 
\be
\widehat{T}_{p, p'} (q)_{\{j \}, \{j'\}} = \prod_{\ell =1}^L \overline{W}_{p,q}(j_\ell - j'_\ell) W_{p',q}(j_\ell - j'_{\ell+1}).
\ele(hTpq)
Then $T_{p,p'} , \widehat{T}_{p, p'}$ commute with the spin-shift operator, denoted again by $X \ (:= \prod_{\ell} X_\ell) $ when no confusion could arise. 
The star-triangle relation (\req(TArel)) yields the following commutative relation
\be
T_{p, p'} (q) \widehat{T}_{p, p'} (r) = (\frac{f_{p'q}f_{pr}}{f_{pq}f_{p'r}})^L T_{p, p'} (r) \widehat{T}_{p, p'} (q) , \ \ \widehat{T}_{p, p'} (q) T_{p, p'} (r) = (\frac{f_{pq}f_{p'r}}{f_{p'q}f_{pr}})^L  \widehat{T}_{p, p'} (r) T_{p, p'} (q) ,
\ele(TTc)
for $q , r \in {\goth W}_{k'}$ (see \cite{BBP} (2.15a)-(2.32b)), by which the $Q$-operators,  defined by 
$$
Q_{p, p'} (q) = \widehat{T}_{p, p'} (q_0)^{-1}  \widehat{T}_{p, p'} (q) = (\frac{f_{pq}f_{p'q_0}}{f_{p'q}f_{p q_0}})^L  T_{p, p'} (q) T_{p, p'} (q_0)^{-1}  
$$
for $q \in {\goth W}_{k'}$, form a commuting family. Here $q_0$ is an arbitrary point in ${\goth W}_{k'}$ for both $\widehat{T}_{p, p'} (q_0), T_{p, p'} (q_0)$ being non-singular.

\subsection{The generalized $\tau^{(2)}$-model \label{ssec.tau2}}
In the discussion of  CPM as a descendent of the six-vertex model, 
a five-parameter family of generalized $\tau^{(2)}$-models was found in \cite{BazS} with the $L$-operator defined by the matrix of $\CZ^2$-auxiliary, $\CZ^N$-quantum space in terms of the Weyl operators $X, Z$:
\be
{\tt L} ( t ) =  \left( \begin{array}{cc}
        1 + t \kappa X  & ( \gamma - \varrho X)Z \\
        t ( \alpha - \beta X)Z^{-1} & t \alpha \gamma + \frac{\beta \varrho}{\kappa} X 
\end{array} \right) , \ \ t \in \CZ .
\ele(Lt2)
Hereafter in this paper, we assume the complex parameters  $\alpha, \beta, \gamma, \varrho, \kappa$ to be non-zero ( even though no such restriction was required in the general discussion  \cite{BazS, GIPS}).  
The above $L$-operator satisfies the Yang-Baxter (YB) equation
\be
R(t/t') ({\tt L} (t) \bigotimes_{aux}1) ( 1
\bigotimes_{aux} {\tt L} (t')) = (1
\bigotimes_{aux} {\tt L}(t'))( {\tt L}(t)
\bigotimes_{aux} 1) R(t/t')
\ele(YB)
for the {\it asymmetric} six-vertex $R$-matrix,
$$
R(t) = \left( \begin{array}{cccc}
        t \omega - 1  & 0 & 0 & 0 \\
        0 &t-1 & \omega  - 1 &  0 \\ 
        0 & t(\omega  - 1) &( t-1)\omega & 0 \\
     0 & 0 &0 & t \omega - 1    
\end{array} \right).
$$
For later convenience, throughout this paper we use another but equivalent labelling of the parameters in (\req(Lt2)): 
$$
({\sf a}, {\sf b}, {\sf a}', {\sf b}', {\sf c} )= (\frac{- \varrho  }{\omega \kappa }, \frac{1}{\gamma}, \frac{\beta }{ \kappa } ,  \frac{-1}{\alpha },   \frac{\kappa}{\alpha \gamma  } ) \in \CZ^{*5} , \  \CZ^* := \CZ \setminus \{ 0 \} ,
$$
i.e. $(\alpha , \beta, \gamma, \varrho , \kappa)= ( \frac{-1}{{\sf b}'}, \frac{-{\sf a}' {\sf c}}{ {\sf b}' {\sf b} }, \frac{1}{\sf b },  \frac{\omega {\sf a c}}{\sf b'b}, \frac{-{\sf c} }{\sf b'b})$; then (\req(Lt2)) becomes 
\be
{\tt L} ( t ) = \left( \begin{array}{cc}
        1  -  t \frac{{\sf c}  }{\sf b' b} X   & (\frac{1}{\sf b }  -\omega   \frac{\sf a c }{\sf b' b} X) Z \\
       - t ( \frac{1}{\sf b'}  -   \frac{\sf a' c}{\sf b' b} X )Z^{-1} & - t \frac{1}{\sf b' b} + \omega   \frac{\sf a' a c }{\sf b' b} X
\end{array} \right) =: \left( \begin{array}{cc}
        {\tt A}(t)  & {\tt B}(t)  \\
        {\tt C}(t)  & {\tt D}(t) 
\end{array} \right).
\ele(Lab')
Note that there is no connection between the above parameters $({\sf a}, {\sf b}, {\sf a}', {\sf b}', {\sf c})$ and the homogeneous coordinates $[a, b, c, d]$ of $\PZ^3$ in (\req(rapidC)). 
The quantum determinant and the "classical" $L$-operator (\cite{GIPS} (88) (45), \cite{R06F} (2.9) (2.24),  \cite{Ta}) of (\req(Lab')) are expressed by 
\bea(ll)
{\rm det}_q {\tt L} (t)  = q (t) X , & q (t):= \frac{\omega {\sf c}}{({\sf b' b})^2} ({\sf a b}- t)({\sf a' b'}- t), \\
{\cal L} (t^N) (: =\langle {\tt L} \rangle ) = \left( \begin{array}{cc}
        \langle A \rangle   & \langle B \rangle \\
        \langle C \rangle &  \langle D \rangle 
\end{array} \right) &= \frac{1}{{\sf b'}^N {\sf b}^N } \left( \begin{array}{cc}
        {\sf b'}^N {\sf b}^N  -  {\sf c}^N t^N  & {\sf b'}^N  - {\sf a}^N {\sf c}^N  \\
        - ( {\sf b}^N - {\sf a'}^N {\sf c}^N )t^N & {\sf a'}^N {\sf a}^N {\sf c}^N - t^N 
\end{array} \right) ,
\elea(qdcL)
where $\langle O \rangle \ := \prod_{i=0}^{N-1} O(\omega^i t)$ denotes the average of the (commuting family of) operators $O(t)$ for $t \in \CZ$. Then the monodromy matrix of the chain size $L$ for (\req(Lab')), 
\be
\bigotimes_{\ell=1}^L  {\tt L}_\ell (t)  =  \left( \begin{array}{cc} A_L(t)  & B_L (t) \\
      C_L (t) & D_L(t)
\end{array} \right), \ \ {\tt L}_\ell (t)= {\tt L}(t) \ {\rm at \ site} \ \ell,
\ele(monM)
again satisfy the YB equation (\req(YB)), with the average given by (\cite{R06F} Proposition 2.2, \cite{Ta}) 
\be 
\langle \bigotimes_{\ell=1}^L  {\tt L}_\ell \rangle := \left( \begin{array}{cc} \langle A_L \rangle  & \langle B_L \rangle \\
      \langle C_L \rangle & \langle D_L\rangle
\end{array} \right) = {\cal L}_1 (t^N) {\cal L}_2 (t^N) \cdots {\cal L}_L (t^N) ( = {\cal L} (t^N)^L). 
\ele(avM)

The $\tau^{(2)}$-matrix is the commuting family of $\stackrel{L}{\otimes} \CZ^N$-operators
defined by the $\omega$-twisted trace of the monodromy matrix (\req(monM)):
\be
\tau^{(2)}(t) = {\rm tr}_{\CZ^2} \bigotimes_{\ell=1}^L  {\tt L}_\ell (\omega t), 
\ele(tau2) 
which commutes with the spin-shift operator $X$. For an integer $j \geq 2$, there exists the $j$th fusion $L$-operator, (a matrix of $\CZ^j$-auxiliary and $\CZ^N$-quantum space) constructed from the $L$-operator (\req(Lab')) by a canonical procedure, which gives rise to the $j$th fusion $\tau^{(j)}$-model, i.e. the commuting $\stackrel{L}{\otimes} \CZ^N$-operators $\tau^{(j)}(t)$ for $t \in \CZ$, so that the operators $\tau^{(j)}$'s  with $\tau^{(0)}=0, \tau^{(1)}= I$ satisfy the following recursive fusion relation:
\bea(l)
\tau^{(2)}(\omega^{j-1} t) \tau^{(j)}(t) =  z( \omega^{j-1} t) X \tau^{(j-1)}(t)  + \tau^{(j+1)}(t) , \ \ j \geq 1 .
\elea(fus)
where $z(t) = q(t)^L$ with $q(t)$ in (\req(qdcL)) (see, e.g. \cite{R06F} Proposition 2.1).

The $\tau^{(2)}$-matrix of CPM with vertical rapidities $p, p'$ in (\req(xymu)), denoted by $\tau^{(2)}_{p, p'}$ hereafter in this paper, is constructed from the $L$-operator (\req(Lab')) with parameters (\req(Lab')) defined by
\be
({\sf a, b, a', b', c}) = (x_p, y_p, x_{p'}, y_{p'}, \mu_p \mu_{p'}) ,
\ele(tpp')
(by formulas (3.37) (3.38) for $j=2, \alpha =n = 0, 1 , m=0, 1$,  (A3), and (3.44a) for $j=2, k = 0$ in \cite{BBP}), and it relates to the CPM transfer matrices (\req(Tpq))(\req(hTpq)) by the $\tau^{(2)}T$-relations (\cite{BBP} (4.20) (4.21))\footnote{An equivalent formulation of first $\tau^{(2)}T$-relation in (\req(tauT)) is given by \cite{BBP} (4.31) using the automorphism $U$ only: 
\be
\tau^{(2)}_{p, p'}(\omega^{-1} t_q) T_{p, p'}(q) = \{\frac{(y_p-  x_q)(t_{p'}-\omega^{-1} t_q) }{y_p y_{p'}(x_{p'}- \omega^{-1} x_q)}\}^L T_{p ,p'}(U^{-1} q) +  \{\frac{\omega \mu_{p'} \mu_p(t_p- t_q)(x_{p'}- x_q) }{y_p y_{p'}(y_p- \omega x_q)}\}^L X T_{p, p'}(U q) .
\ele(tauT')
}:
\bea(l)
\tau^{(2)}_{p, p'}(\omega^{-1} t_q) T_{p, p'}(q) = \{\frac{(y_p-  x_q)(t_{p'}-\omega^{-1} t_q) }{y_p y_{p'}(x_{p'}- \omega^{-1} x_q)}\}^L T_{p ,p'}(U^{-1} q) + \{\frac{(y_{p'}- y_q)(t_p- t_q) }{y_p y_{p'}(x_p-y_q)}\}^L T_{p, p'}(R^2 U^{-1}  q) , \\
\widehat{T}_{p, p'}(q) \tau^{(2)}_{p, p'}(\omega^{-1} t_q)  = \{\frac{(y_{p'}- x_q)(t_p-\omega^{-1} t_q) }{y_p y_{p'}(x_p-\omega^{-1} x_q)}\}^L \widehat{T}_{p, p'}(U^{-1} q) + \{\frac{(y_p- y_q)(t_{p'}- t_q) }{y_p y_{p'}(x_{p'}-y_q)}\}^L \widehat{T}_{p, p'}(R^2U^{-1} q) ,
\elea(tauT)
where $U, R$ are ${\goth W}_{k'}$-automorphisms defined in (\req(Aut)). Then $\tau^{(2)}_{p, p'}$'s form a $3$-parameter family among all the generalized $\tau^{(2)}$-matrices. On the other hand, there are two equivalent relations among all $L$-operator (\req(Lab')) which produce equivalent $\tau^{(2)}$-models. First, the $\tau^{(2)}$-matrix is unchanged when applying the gauge transform to the $L$-operator (\req(Lab')) by $M L(t) M^{-1} $ with $M = {\rm dia}[1, \nu]$. The corresponding change of parameters is the transformation:
\be
({\sf a, b, a', b', c}) \mapsto (\nu^{-1} {\sf a}, \nu {\sf b}, \nu {\sf a}', \nu^{-1} {\sf b}', {\sf c} ), \ \ \nu \in \CZ^*.
\ele(nu)
The second equivalent relation is induced by substituting the variable $t$ by $\lambda^{-1}  t$, which corresponds to the transformation of parameters:
\be
({\sf a, b, a', b', c})  \mapsto (\lambda {\sf a}, {\sf b}, {\sf a}', \lambda {\sf b}', {\sf c} ), \ \ \lambda \in \CZ^*.
\ele(lambda)
Then the relations, (\req(nu)) and (\req(lambda)), give rise to a $\CZ^{* 2}$- action of the $5$-parameters of $L$-operators (\req(Lab')), by which a generic $\tau^{(2)}$-model can be reduced to a CPM $\tau^{(2)}_{p, p'}$ in (\req(tpp')). Indeed, an explicit description of  $\tau^{(2)}$-models equivalent to the chiral Potts $\tau^{(2)}_{p, p'}$ is described by the following theorem, the proof of which  we leave in the appendix.
\begin{thm} \label{thm:CPt}
The necessary and sufficient condition for parameters ${\sf a, b, a', b', c}  \in \CZ^*$ in $(\req(Lab'))$ whose  $\tau^{(2)}$-model is equivalent to a CPM $\tau^{(2)}_{p, p'}$-model via relations $(\req(nu)), (\req(lambda))$ for $p, p'$ in  $(\req(xymu))$ with $k' \neq 0, \pm 1$  is 
\be
\left\{ \begin{array}{ll}  {\sf b}^N -{\sf a}'^N = {\sf b}'^N -{\sf a}^N= 0  & \mbox{for ${\sf c}^N =1$ } \\
({\sf b}^N -{\sf a}'^N)({\sf b}'^N -{\sf a}^N )({\sf b}^N -{\sf a}'^N{\sf c}^N)({\sf b}'^N - {\sf a}^N {\sf c}^N)({\sf b}^N {\sf b}'^N - {\sf a}^N {\sf a}'^N {\sf c}^N) \neq 0 & \mbox{for  ${\sf c}^N \neq 1$ }. \end{array} \right.
\ele(tCPM)
In the above situation, the $\tau^{(2)}$-matrix  is related to $\tau^{(2)}_{p, p'}$-matrix by a change of variables, $t= \lambda^{-1} \widetilde{t}$ for some $\lambda \in \CZ^*$:
\be
\tau^{(2)}(t) = \tau^{(2)}_{p, p'} ( \widetilde{t}).
\ele(tt)
\end{thm}
\par \vspace{.1mm}
As an easy consequence of the above theorem, a conjectural  boundary fusion relation (\cite{GIPS} (107), \cite{R06F} (2.30)) is valid for a generalized $\tau^{(2)}$-model: 
\begin{cor} \label{cor:fuT}
the boundary fusion relation holds for an arbitrary $\tau^{(2)}$-model:
\be
\tau^{(N+1)}(t) = z(t) X \tau^{(N-1)}(\omega t) + u(t) I ,
\ele(bfu)
where $z(t)= q(t)^L$ as in $(\req(fus))$, and $u(t):= \langle A_L \rangle + \langle D_L \rangle$ with $A_L , D_L$ in $(\req(monM))$.
\end{cor}
{\it Proof}. By the construction of $\tau^{(j)}$-matrices (\cite{R06F} section 2.2) and using the continuity argument, one needs only to verify the relation (\req(bfu)) for a generic $\tau^{(2)}$-model as described in (\req(tCPM)), which by Theorem \ref{thm:CPt}, is equivalent to a CPM $\tau^{(2)}_{p, p'}$-model. Note that the relation (\req(nu)) leaves the quantum determinant $q(t)$ unchanged, but changes ${\cal L}(t^N)$ in (\req(qdcL))  only by a gauge 
transformation, hence with the same $u(t)$ in (\req(bfu)) by (\req(avM)). For the relation (\req(lambda)), one uses $t = \lambda^{-1} \widetilde{t}$ and $(\lambda {\sf a}, {\sf b}, {\sf a}', \lambda {\sf b}', {\sf c} )= (\widetilde{\sf a}, \widetilde{\sf b}, \widetilde{\sf a}', \widetilde{\sf b}', \widetilde{\sf c} )$, then finds $q(t) = q(\widetilde{t})$, and ${\cal L}(t^N) = {\cal L}(\widetilde{t}^N)$, hence $ u(t)= u(\widetilde{t})$. Therefore the equality (\req(bfu)) is preserved under relations (\req(nu)) and (\req(qdcL)). The conclusion of this theorem now follows from the known fact about the valid boundary fusion relation (\req(bfu)) for CPM $\tau^{(2)}_{p, p'}$-model (\cite{BBP} (4.27c) (4.28) (4.29) (2.46)), where $q(t)= \frac{\omega \mu_p \mu_{p'}(t_p-t)(t_{p'}-t)}{y_p^2 y_{p'}^2}$, $u(t)= \alpha_q + \overline{\alpha}_q$ with $\alpha_q= e_q^L,  \overline{\alpha}_q = \overline{e}_q^L$ and $e_q, \overline{e}_q$ the eigenvalues of ${\cal L}(t^N)$ in (\req(qdcL)):
\bea(llll)
\alpha_q= e_q^L, &e_q = \frac{\mu_q^N(y_p^N-x_q^N)(y_{p'}^N-x_q^N)}{k' y_p^N y_{p'}^N} & = \frac{(t_p^N-t_q^N)(y_{p'}^N-x_q^N)}{(x_p^N-x_q^N)y_p^N y_{p'}^N} & = \frac{(y_p^N-x_q^N)(t_{p'}^N-t_q^N)}{(x_{p'}^N-x_q^N)y_p^N y_{p'}^N} , \\ 
 \overline{\alpha}_q = \overline{e}_q^L, & \overline{e}_q = \frac{\mu_q^{-N}(y_p^N-y_q^N)(y_{p'}^N-y_q^N)}{k' y_p^N y_{p'}^N} & = \frac{(y_p^N-y_q^N)(t_{p'}^N-t_q^N)}{(x_{p'}^N-y_q^N)y_p^N y_{p'}^N} & = \frac{(t_p^N-t_q^N)(y_{p'}^N-y_q^N)}{(x_p^N-y_q^N)y_p^N y_{p'}^N} .
\elea(eeq)
\vspace{.1in} \par  \noindent
{\bf Remark} The functions, $\alpha_q, \overline{\alpha}_q$ in (\req(eeq)) and $z(t)$ in (\req(bfu)), satisfy 
$\alpha_q \overline{\alpha}_q = z(t)z(\omega t) \cdots z(\omega^{N-1}t)$, which is the relation between the determinant of (\req(avM)) and the quantum determinant of monodromy matrix (\req(monM)):  $
{\rm det}  \langle \otimes_\ell {\tt L}_\ell \rangle =  \langle {\rm det}_q \otimes_\ell {\tt L}_\ell \rangle$.
\vspace{.2in} \par  
Note that each condition in (\req(tCPM)) is preserved under $(\req(nu))$ and $(\req(lambda))$, and with the same criterion to which both $\tau^{(2)}_{p, p'}$ and $\tau^{(2)}$ in (\req(tt)) belong. The structure of $\tau^{(2)}$-model for the case ${\sf c}^N =1$ in (\req(tCPM)) can be determined as follows. By Lemma \ref{thm:CPt}, we need only to consider those CPM $\tau^{(2)}_{p, p'}$ with $(\mu_p \mu_{p'})^N=1$, then the relation (\req(xymu)) yields 
$(x_{p'}, y_{p'}, \mu_{p'})= ( \omega^i y_p, \omega^{-j} x_p, \omega^k \mu_p^{-1} )$ for some $i, j, k \in \ZZ_N$. The $L$-operator (\req(Lab')) under the gauge transform (\req(nu)) with $\nu = y_p^{-1}$ now becomes
$$
{\tt L} ( {\tt t} ) =  
 \left( \begin{array}{cc}
        1 - {\tt t} \omega^k X  & ( 1 -  \omega^{j+k+1} X)Z \\
        -  {\tt t} ( 1  -  \omega^{i+k}  X)Z^{-1} & -{\tt t} + \omega^{i+j+k+1}  X 
\end{array} \right) 
$$
where ${\tt t}= \frac{\omega^j  t}{ x_py_p } $. Note the above ${\tt L}(t)$ is the  $L$-operator (\req(Lab')) with ${\sf b,  b'}=1$ and $({\sf a,  a', c})=(\omega^j, \omega^i, \omega^k)$.  In particular, it is represented by the $L$-operator of the superintegrable $\tau^{(2)}_{p, p'}$ with $(x_p, y_p, \mu_p) = (\omega^m \eta^{\frac{1}{2}}, \omega^{m'} \eta^{\frac{1}{2}}, \omega^n )$ , $(x_{p'}, y_{p'}, \mu_{p'}) = (\omega^{m'+i} \eta^{\frac{1}{2}}, \omega^{m-j} \eta^{\frac{1}{2}}, \omega^{-n+k} )$ where $\eta:= (\frac{1-k'}{1+k'})^\frac{1}{N}$, among which the homogenous CPM $\tau^{(2)}$-matrices are those with the relations, $\omega^i = \omega^j= \omega^{m-m'},  \omega^k = \omega^{2n}$. When $N$ is odd, those among superintegrable $\tau^{(2)}_{p, p'}$  
with $\omega^{i+j+1}= \omega^k=1$ correspond to  XXZ chains for the cyclic $U_q(sl_2)$ representations with $q^N=1$ and a $N$th root-of-unity representation parameter $\varsigma^N =1$  \cite{R075}.

 \section{ The Q-operator of the generalized $\tau^{(2)}$-model \label{sec:Qtau} }
\setcounter{equation}{0}
This section is devoted to the construction of $Q$-operator  for a generalized $\tau^{(2)}$-model with the $L$-operator (\req(Lab')). The $\tau^{(2)}$-models of our main interest are  those not equivalent to CPM $\tau^{(2)}$-models in (\req(tCPM)). By following Baxter's method of producing the eight-vertex $Q_{72}$-operator in \cite{B72}, we shall first  in subsection \ref{ssec.QtCP} describe the general mechanism  of constructing the $Q$-operator of a generalized $\tau^{(2)}$-model, and illustrate the method in the CPM $\tau^{(2)}$-model by  reproducing the transfer matrices (\req(Tpq)), (\req(hTpq)) as the $Q_R$ and $Q_L$-operators. Then in subsection \ref{ssec.sdCP},
we identify the transfer matrix of the selfdual degenerate Potts models with $k'=1$ as the $Q$-operator for $\tau^{(2)}$-models with ${\sf c}^N=1$, among which when $N$ is odd, are those equivalent to XXZ chains associated to  cyclic $U_q(sl_2)$ representations with $q^N=1$ and the representation parameter $\varsigma^N \neq 1$  \cite{R075}.

\subsection{Construction of the $Q$-operator of a generalized $\tau^{(2)}$-model, and the transfer matrix of the chiral Potts model  \label{ssec.QtCP}}
For a $L$-operator (\req(Lab')), we define the $\CZ^N$-operators ${\tt A}_\eta, {\tt C}_{\xi, \eta}, {\tt D}_\xi$ for $\xi, \eta \in \CZ$ as in \cite{R075},
\bea(l)
{\tt A}_\eta (t):= {\tt A}(t)  -  {\tt B}(t) \eta,  ~ ~ ~  {\tt D}_\xi(t):= \xi {\tt B}(t)   + {\tt D}(t)  ,   \\
{\tt C}_{\xi, \eta} (t) :=  \xi {\tt A}(t)  + {\tt C}(t) - \xi  {\tt B}(t)\eta  -{\tt D}(t)\eta ,
\elea(gauL)
with the following commutative relations (\cite{R075} (3.5)):
\bea(ll)
{\tt C}_{\xi, \eta} (t) X^{-1} {\tt A}_\eta(\omega t) =  {\tt A}_\eta (t)X^{-1}{\tt C}_{\xi, \eta}(\omega t), &
{\tt C}_{\xi, \eta} (\omega t) {\tt D}_\xi(t) =  {\tt D}_\xi (\omega t) {\tt C}_{\xi, \eta}(t) .
\elea(ADCg)
We are going to follow Baxter's $Q_{72}$-operator method in \cite{B72} to produce the $Q_R$, $Q_L$, and $Q$-operator associated to the $L$-operator  (\req(Lab')).  The $Q_R$, $Q_L$-matrices are  defined by 
\be
Q_R= {\rm tr}_{\CZ^N} ( \bigotimes_{\ell =1}^L {\sf S}_{\ell}), \ Q_L= {\rm tr}_{\CZ^N} ( \bigotimes_{\ell =1}^L \widehat{\sf S}_{\ell}) 
\ele(QRL)
with ${\sf S}_{\ell}$, $\widehat{\sf S}_{\ell}$ = ${\sf S}, \widehat{\sf S}$  at site $\ell$ respectively, where the local ${\sf S }, \widehat{\sf S }$-operators are matrices of $\CZ^N$-auxiliary and $\CZ^N$-quantum space with the $\CZ^N$-operator-entries ${\sf S }_{i,j}, \widehat{\sf S}_{i, j}$:
\be
{\sf S }  = ({\sf S }_{i,j})_{i, j \in \ZZ_N}  , \ \widehat{\sf S} = ( \widehat{\sf S}_{i, j})_{i, j \in \ZZ_N} .
\ele(SSmat)
Consider the local-operator ${\sf U}$ with the $\CZ^2 \otimes \CZ^N $-auxiliary and $\CZ^N$-quantum space:
$$
{\sf U} = \left( \begin{array}{cc}
        {\sf A} {\sf S } & {\sf B} {\sf S} \\
        {\sf C} {\sf S} & {\sf D} {\sf S }
\end{array} \right) .
$$
Hereafter we write the operators ${\sf A} (t), {\sf B} (t), {\sf C}(t), {\sf D}(t) $ simply by ${\sf A}, {\sf B}, {\sf C}, {\sf D}$ if no confusion could arise; while the matrix ${\sf S}$ will depend on some variable $\sigma$ algebraically related to the variable $t$: ${\sf S} = {\sf S}(\sigma)$. Then one has $\tau^{(2)} Q_R = {\rm tr}_{\CZ^2 \otimes \CZ^N} ( \bigotimes_{\ell =1}^L {\sf U}_{\ell})$ where ${\sf U}_{\ell}= {\sf U}$ at the site $\ell$, and $\tau^{(2)} Q_R$ will be decomposed into the sum of two matrices if we can find a $2N$ by $2N$ scalar matrix (independent of $\sigma$) 
\be
{\sf M} = \left( \begin{array}{cc}
        I_N  & 0 \\
       -\delta & I_N 
\end{array} \right) , \ \ \delta = {\rm dia} [\delta_0, \cdots, \delta_{N-1}] ,
\ele(Md)
so that the matrix 
$$
{\tt M}^{-1} {\tt U} {\tt M} = \left( \begin{array}{cc}
        {\tt A}_{\delta_j}{\tt S }_{i,j}  ,  & {\tt B} {\tt S }_{i,j}   \\ 
       {\tt C}_{\delta_i, \delta_{j}}{\tt S }_{i,j}  , & {\tt D}_{\delta_i} {\tt S }_{i,j} 
\end{array} \right)_{i, j \in \ZZ_N} 
$$
has vanishing lower blocktriangular matrix; and a similar discussion also for $ Q_L \tau^{(2)}$. For this purpose, we first determine the condition of $\xi, \eta$ so that the ${\tt C}_{\xi, \eta}$ in (\req(gauL)) is a singular operator.
Since the entries of ${\tt C}_{\xi, \eta}$ are zeros except
$$
{\sf b b'} \langle  n | {\tt C}_{\xi, \eta}| n \rangle = ({\sf b b'} \xi - \omega^{-n}{\sf b} t ) (1 - \omega^n {\sf b}^{-1} \eta ) , ~ \ ~  {\sf b b'} \langle  n | {\tt C}_{\xi, \eta}| n-1 \rangle = -{\sf c} ( \xi - \omega^{-n+1}{\sf a'} ) ( t - \omega^n  \eta {\sf a}) , 
$$
one finds
\be
({\sf b b'})^N {\rm det} ~ {\tt C}_{\xi, \eta} = ({\sf b'}^N \xi^N -  t^N ) ({\sf b}^N -  \eta^N )- {\sf c}^N ( \xi^N - {\sf a'}^N ) ( t^N - \eta^N  {\sf a}^N ). 
\ele(detC)
The vanishing determinant of ${\tt C}_{\xi, \eta}$ will provide the criterion of $\xi, \eta$ with a non-zero kernel vector of ${\tt C}_{\xi, \eta}$, by the same argument as in Lemma 3.2 of \cite{R075}, now explained below. If ${\rm det}~ {\tt C}_{\xi, \eta}=0$, the kernel of ${\tt C}_{\xi, \eta}$ is one-dimensional generated by the kernel vector $v = \sum_{n \in \ZZ_N} v_n | n \rangle$ defined by
\be
\frac{v_n}{v_{n-1}}   = \frac{{\sf c} ( \omega {\sf a'} - \omega^n \xi  ) ( t - \omega^n  {\sf a} \eta )}{ ({\sf b} - \omega^n  \eta ) (  t  - \omega^n {\sf b'} \xi )}.
\ele(kv)
Hence by (\req(ADCg)), one finds the relations
\be
{\tt A}_\eta( t)v(t) = \lambda(t)X v(\omega^{-1} t) , \ ~ \ ~ {\tt D}_\xi(t)v(t) = \lambda'(t) v (\omega t) 
\ele(ADv)
where $\lambda(t) =  \frac{ {\tt c}  (\omega {\sf a'} {\sf b'}  -    t ) (  t -   \omega {\sf a} \eta ) v_0(t)}{ {\sf bb'} (  t  -  \omega {\sf b'} \xi ) v_0(\omega^{-1} t) } $, $   \lambda'(t)
= \frac{ ( {\sf a}{\sf b} -  t ) (  t  -  {\sf b'} \xi ) v_0(t)}{{\sf bb'} ( t -   {\sf a} \eta )v (\omega t)_0 } $.  
A similar argument implies that the one-dimensional cokernel  for a singular operator ${\tt C}(\xi^*, \eta^*)$ is generated by  $v^* = \sum_{n \in \ZZ_N} v^{*n} \langle n |$ with
\be
 \frac{v^{* n}}{v^{* n-1}}  = \frac{ ({\sf b} - \omega^{n-1}  \eta^* )( t - \omega^{n-1} {\sf b'} \xi^*  )}{ {\sf c} ( {\sf a'}  - \omega^{n-1} \xi^* ) ( t - \omega^n {\sf a} \eta^* )} ,
\ele(ckv)
satisfying the relations 
\be
v^* (t) {\tt A}(\eta^*) (t) = \lambda^*(t)v^* (\omega t)X  , \ ~ \ ~  v^* ( t)  {\tt D}(\xi^*) ( t) = \lambda^{* '} (t) v^*(\omega^{-1} t) 
\ele(ADv*)
where $\lambda^*(t)= \frac{ {\sf c} ({\sf a'}{\sf b'}-t) (t - {\sf a} \eta^* )v^{*0} (t)}{ {\sf bb'} ( t - \omega ^{-1}   {\sf b'} \xi^*  ) v^{* 0} (\omega t) }
 $ , $\lambda^{* '}(t)  =  \frac{ ( \omega {\sf a} {\sf b} - t ) ( t -   {\sf b'} \xi^*  ) v^{*0} ( t) }{ {\sf  bb'} ( t - \omega {\sf a} \eta^* )v^{*0}(\omega^{-1} t) } $. 
In the above discussion when applying to the CPM $L$-operator with parameters in (\req(Lab')) given by (\req(tpp')) with $t= t_q$, we set the parameters in (\req(kv)) (\req(ckv))  by
\be 
\xi = \omega^{-i} x_q, \eta= \omega^{-j} x_q ; \ \ \xi^* = \omega^{-i} y_q, \eta^*= \omega^{-j} y_q ,
\ele(xixi*) 
so that the cyclic vectors are determined by the relations 
\bea(ll)
\frac{v_n}{v_{n-1}}   = \frac{\mu_p \mu_{p'} ( \omega x_{p'} - \omega^{n -i} x_q  ) ( y_q - \omega^{n-j}  x_p )}{ ( y_q - \omega^{n-i} y_{p'} ) (y_p - \omega^{n-j}   x_q )}, & 
 \frac{v^{* n}}{v^{* n-1}}  = \frac{( x_q - \omega^{n-1-i} y_{p'}  ) (y_p - \omega^{n-1-j} y_q )}{\mu_p \mu_{p'}  ( x_{p'}  - \omega^{n-1-i} y_q ) ( x_q - \omega^n {\sf a} \omega^{-j} )}.
\elea(cpmv)
With the following functions $\lambda(t), \lambda'(t)$ in  (\req(ADv)), we define the vectors in  (\req(kv)) by\footnote{Using a different value for $v_0(\omega t)$, one may define the vector $v_n(\omega t)$ in (\req(CPMv)) by $v_n (\omega t; \omega^{-i} x_q, \omega^{-j} x_q)  = \overline{W}_{p', U q}(n-i-1)W_{p, U q}(n-j-1)$, hence change the value of $\lambda'(t)$ by 
$\lambda'(t) = \frac{\omega \mu_{p'} \mu_p ( t_p -t)  (x_{p'}-x_q)  }{y_p y_{p'}  (y_q- \omega x_p)}$, which provides another form (\req(tauT')) in the $\tau^{(2)}T$-relation.}
\bea(ll)
v_n (t; \omega^{-i} x_q, \omega^{-j} x_q) &= \overline{W}_{p', q}(n-i)W_{p, q}(n-j), \\ 
v_n(\omega^{-1} t; \omega^{-i} x_q, \omega^{-j} x_q) &= \overline{W}_{p', U^{-1} q}(n-i+1)W_{p, U^{-1} q}(n-j+1), \\ 
v_n (\omega t; \omega^{-i} x_q, \omega^{-j} x_q)  &= \overline{W}_{p', R^2U^{-1} q}(n-i)W_{p, R^2U^{-1} q}(n-j) ; \\
\lambda(t) =  \frac{  ( t_{p'}  - \omega^{-1}   t )  (y_p -  x_q)}{ y_p y_{p'} 
(  x_{p'}- \omega^{-1} x_q) } , & \lambda'(t) = \frac{\omega^{-i+j} ( t_p -  t )  (y_q- y_{p'})  }{y_p y_{p'}  (y_q- x_p)},
\elea(CPMv)
where $\overline{W}_{p', q}, W_{p, q} $ are Boltzmann weights in (\req(CPW)), and the automorphisms  $U, R$ are in (\req(Aut)).  
Similarly, the cyclic vectors in (\req(ckv))  and functions in (\req(ADv*)) are expressed by
\bea(ll)
v^{*n} (t; \omega^{-i} y_q, \omega^{-j} y_q) &= W_{p', q}(i-n)\overline{W}_{p, q}(j-n), \\ 
v^{*n} (\omega^{-1} t; \omega^{-i} y_q, \omega^{-j} y_q) &= W_{p', R^2U^{-1}q}(i-n+1)\overline{W}_{p, R^2U^{-1}q}(j-n+1) , \\ 
v^{*n} (\omega t; \omega^{-i} y_q, \omega^{-j} y_q)  &= W_{p', U^{-1} q}(i-n)\overline{W}_{p, U^{-1} q}(j-n) , \\
\lambda^*(t)= \frac{  (t_{p'}-t)  (y_p - y_q )  }{ y_p y_{p'} (x_{p'} - y_q) }, & \lambda^{* '}(t) =  \frac{\omega^{-i+j} ( t_p - \omega ^{-1} t ) (y_{p'} - x_q)  }{ y_p y_{p'} (x_p - \omega^{-1} x_q)  } .
\elea(CPMv*)
(Note that when the rapidities $p, p'$ are superintegrable elements, the cyclic vectors and functions in (\req(CPMv)) (\req(CPMv*)) were derived as formulas (4.15), (4.16) in \cite{R075}). We now construct 
the $Q_R$, $Q_L$-operators (\req(QRL)) for two arbitrary elements $p, p' \in {\goth W}_{k'}$ using the following ${\sf S}, \widehat{\sf S}$-matrices (\req(SSmat)) as defined in \cite{R075} (3.36): 
\be
{\sf S}_{i,j} = v ( t_q; \omega^{-i} x_q, \omega^{-j} x_q) \langle j | , ~ ~  \widehat{\sf S}_{i, j} = | j \rangle v^{*} (t_q; \omega^{-i} y_q, \omega^{-j} y_q) ,
\ele(SSij)
which in turn yields the identification of $Q_R$, $Q_L$-operators with the CPM transfer matrices in (\req(Tpq)) (\req(hTpq)):
\be
Q_R (q) = T_{p, p'}(q) , \ ~ \ ~ Q_L (q) = \widehat{T}_{p, p'}(q)
\ele(QTpp)
for $q \in {\goth W}_{k'}$. Then the $\tau^{(2)}T$-relation (\req(tauT)) follows from (\req(CPMv)) (\req(CPMv*)).

We now use the kernel vector of the ${\tt C}_{\xi, \eta}$-operator to construct the $Q_R$-operator of an arbitrary $\tau^{(2)}$-model with the $L$-operator in (\req(Lab')). Set $(\xi, \eta)= (\delta_i, \delta_j)$ with $\delta_i$'s in (\req(Md)), and the operator ${\sf S }_{i,j}$ in (\req(SSmat)) by ${\tt S}_{i, j} = {\tt v}_{i, j} \tau_{i, j}$,
where ${\tt v}_{i, j}$ is the cyclic kernel vector in ${\tt C}_{\delta_i, \delta_j}$, and $\tau_{i, j} \in \CZ^{N *}$ is a parameter vector. The vector ${\tt v}_{i, j}$ is non-zero if ${\tt C}_{\xi, \eta}$ is a singular matrix. By (\req(detC)), one finds
$$
({\sf b b'})^N ( {\rm det}~{\tt C}_{\xi, \eta} - {\rm det}~{\tt C}_{\eta, \xi}) =
(\xi^N - \eta^N) ( {\sf b}^N {\sf b'}^N+{\sf c}^N  {\sf a}^N {\sf a'}^N - t^N (1+ {\sf c}^N)).
$$
Therefore it is convenient to assume 
\be
\xi^N = \eta^N ,
\ele(xi=e)
i.e. all $\delta_i^N$ with the same value $\xi^N$ so that ${\tt C}_{\delta_i, \delta_j}$ are singular matrices for all $i, j$, which by (\req(detC)) are equivalent to the relation
\be
({\sf b}^N {\sf b'}^N- {\sf c}^N{\sf a}^N{\sf a'}^N) \xi^N   + ({\sf c}^N {\sf a'}^N- {\sf b}^N) t^N +  ({\sf c}^N{\sf a}^N -  {\sf b'}^N) \xi^{2N} +  (1-{\sf c}^N) \xi^N t^N  = 0. 
\ele(dxi)
We shall conduct the $Q$-operator investigation under the above assumption, with our main interest especially on those $L$-operator (\req(Lab')) not equivalent to CPM ones as described in Theorem \ref{thm:CPt}. Note that the variable $\xi$ in (\req(dxi)) is algebraically related to $t$ except the case when ${\sf c}^N-1= {\sf a}^N -  {\sf b'}^N=0$, equivalent to the first relation in (\req(tCPM)), which was previously discussed in \cite{R075} with {\it no constraint} on $\xi^N$. Furthermore, in the case for the CPM $\tau^{(2)}$-model with parameters in (\req(tpp')), one finds 
$$
{\sf b}^N {\sf b'}^N- {\sf c}^N{\sf a}^N{\sf a'}^N= 1-{\sf c}^N = -k(
{\sf c}^N {\sf a'}^N- {\sf b}^N)= -k({\sf c}^N{\sf a}^N -  {\sf b'}^N),
$$
which is equal to $1-\mu_p^N \mu_{p'}^N $. Then by using (\req(xixi*)), (\req(dxi)) becomes the first rapidity relation in (\req(xyx)). Hence one may regard the relation  (\req(dxi)) as the rapidity-constraint for a generalized $\tau^{(2)}$-model as it will become clearer later in the paper.

First we consider the case
\be
{\sf c}^N {\sf a'}^N- {\sf b}^N= 0 ~ \ {\rm or} \ ~ ~  {\sf c}^N {\sf a}^N -{\sf b'}^N = 0. 
\ele(aba)
When ${\sf c}^N {\sf a'}^N  = {\sf b}^N$, $\xi=0$ is a solution of (\req(dxi)), equivalently to say, the entry ${\tt C}(t)$ of the $L$-operator (\req(Lab')) possesses a non-zero kernel vector; similarly, there exists a non-zero kernel vector of ${\tt B}(t)$-matrix when ${\sf c}^N{\sf a}^N =  {\sf b'}^N$. Such a kernel vector defines the pseudo-vacuum state in the algebraic Bethe ansatz method, by which the eigenvalue problem was previously investigated in \cite{R06F}. Hence for the rest of this paper, we shall restrict our discussion only on the remaining cases, i.e. with the condition
\be
({\sf c}^N {\sf a'}^N- {\sf b}^N)( {\sf c}^N {\sf a}^N -{\sf b'}^N) \neq 0. 
\ele(nz)
In the next subsection we construct the $Q$-operator of the $\tau^{(2)}$-matrix with (\req(nz)) and ${\sf c}^N=1$ through the selfdual Potts models as degenerate forms of CPM.

\subsection{Selfdual degenerate Potts models \label{ssec.sdCP}}
In this subsection, we consider the $L$-operator (\req(Lab')) with the condition (\req(nz)) and ${\sf c}^N=1$. First, we assume ${\sf b}^N {\sf b'}^N = {\sf a}^N{\sf a'}^N$. Using the relation (\req(nu)) or (\req(lambda)), one may reduce the case with the condition
\be
 {\sf a}^N + {\sf b}^N = {\sf a'}^N + {\sf b'}^N = 0 ,
\ele(ab0)
for $ {\sf a}^N \neq {\sf b'}^N $,
hence ${\sf a'}^N- {\sf b}^N= {\sf a}^N -  {\sf b'}^N \neq 0$ by (\req(nz)). Introduce the variables $x, y$ with $x^N= \xi^N$ and $xy=t$, then the equation (\req(dxi)) becomes
$x^N + y^N  = 0$, 
which can be regarded as the degenerate rapidities in (\req(rapidC)) for $k=0, k' = \pm 1$ with elements expressed by 
\be
q: (x_q, y_q, \mu_q) , ~ ~  \ x_q^N + y_q^N = 0, \ \mu_q^N = \pm 1 .
\ele(xy0)
The Boltzmann weights (\req(CPW)) with $p, q$ in the curve (\req(xy0)) provide the  
selfdual solution of the star-triangle equation (\req(TArel)) ((\cite{BPA} (10) for $I=0$, \cite{FatZ})). Define the elements $p, p'$ in (\req(xy0)) by 
\be
p: (x_p, y_p, \mu_p) = ({\sf a}, {\sf b}, 1) , \ \ p': (x_{p'}, y_{p'}, \mu_{p'}) = ({\sf a'}, {\sf b'}, {\sf c}). 
\ele(pp1c)
Then (\req(tpp')) holds, and we obtain the relation (\req(cpmv)) with parameters given by (\req(xixi*)). The ${\sf S}$-matrices (\req(SSmat)) defined by formulas (\req(CPMv)) (\req(CPMv*)) (\req(SSij)) give rise to the $Q_R$, $Q_L$-operators by using the identity (\req(QTpp)), where  $T_{p, p'}(q), \widehat{T}_{p, p'}(q)$ are the transfer matrices (\req(Tpq)), (\req(hTpq))  of
the selfdual Potts model with the rapidity $q$ in (\req(xy0)). Then $\tau^{(2)}T$-relation (\req(tauT)) holds. Note 
that when $N$ is odd, for a $N$th root-of-unity ${\sf q}$  the XXZ chains associated to cyclic representations of $U_{\sf q}(sl_2)$ are known \cite{R075} to be equivalent to the $\tau^{(2)}$-models with the $L$-operator  (\req(Lab')) satisfying the conditions: ${\sf b, b', c}=1, \omega  {\sf a} {\sf a'}= 1$, where ${\sf a'}=: \varsigma $ is the parameter of  $U_{\sf q}(sl_2)$-cyclic representations. 
By Theorem \ref{thm:CPt}, the CPM $\tau^{(2)}$-model occurs only when $\varsigma^N=1$, in which case the $Q$-operator is equal to the CPM transfer matrix with two vertical superintegrable rapidities (\cite{R075} Theorem 4.2, or section \ref{ssec.tau2} of this paper). 
The $L$-operator ${\tt L} ( {\tt t} )$ in (\req(Lab')) for the rest cases with  $({\sf a, b, a', b', c})= (\omega^{-1} \varsigma^{-1}, 1, \varsigma, 1, 1)$~$(\varsigma^N \neq 1)$ satisfy the conditions in our previous discussion: ${\sf c}^N=1, {\sf b}^N {\sf b'}^N = {\sf a}^N{\sf a'}^N$ and (\req(nz)), to which the theory of selfdual degenerate CPM can be applied. Indeed by the above discussion, the parameter in (\req(ab0)) and the variable $t_q =x_q y_q$ in (\req(xy0)) are related to $\varsigma$ and ${\tt t}$  in the following manner when using the relation (\req(nu)): 
$$
({\sf a, b, a', b', c}) = ( \omega^{-1} {\rm i}^\frac{-1}{N} \varsigma^\frac{-1}{2} , {\rm i}^\frac{1}{N} \varsigma^\frac{-1}{2} , {\rm i}^\frac{1}{N} \varsigma^\frac{1}{2} , {\rm i}^\frac{-1}{N} \varsigma^\frac{1}{2} , 1), \ ~ ~  t_q = {\tt t}, \ \ ({\rm i}= \sqrt{-1}).  
$$
or by using the relation (\req(lambda)) with
$$
({\sf a, b, a', b', c}) = (\sqrt[N]{-1} \omega^{-1}, 1, \varsigma, \sqrt[N]{-1} \varsigma, 1), \ ~ ~  t_q = \sqrt[N]{-1}  \varsigma  {\tt t}. 
$$

We now consider the case ${\sf b}^N {\sf b'}^N \neq {\sf a}^N{\sf a'}^N$ with ${\sf c}^N = 1$ and the condition (\req(nz)).
Using relations (\req(nu)) and (\req(lambda)), one may assume
$$
{\sf a}^N{\sf a'}^N -{\sf b}^N {\sf b'}^N = {\sf a'}^N- {\sf b}^N= {\sf a}^N -  {\sf b'}^N \neq 0 ,
$$
which is equivalent to 
\be
{\sf a}^N +  {\sf b}^N = {\sf a'}^N + {\sf b'}^N  = 1 ,
\ele(ab2)
with $ {\sf a}^N \neq  {\sf b'}^N$.
The coordinates $(x, y)$ with $x^N= \xi^N$ and $xy=t$ in (\req(dxi)) in turn yields the equation of the Fermat curve, 
$$
x^N +  y^N  = 1,
$$ 
which can be realized as a degenerated form of the chiral Potts curve (\req(rapidC)) in $\PZ^3$ with $k'=1$ (by rescaling a factor on $c, d$-components): 
\be
k'=1 : \  a^N + b^N = 1, \ c^N = d^N = 1   \ \Longleftrightarrow \ x^N + y^N = 1, \ \mu^N=1 .
\ele(xy1)
The Boltzmann weights (\req(CPW)) with rapidities in (\req(xy1)) define  a selfdual solution  of (\req(TArel)) (\cite{BPA} (10) for $I=1$, \cite{MPTS, AMPT}).  With $p, p'$ defined in (\req(pp1c)) and $q$ in (\req(xy1)), the relations 
(\req(xixi*))-(\req(QTpp)) are valid, and the transfer matrices $T_{p, p'}(q), \widehat{T}_{p, p'}(q)$ of the selfdual degenerate CPM associated to the (\req(xy1)) give rise to the $Q_R$, $Q_L$-operators with the commutation relation (\req(TTc)) and $\tau^{(2)}T$-relation (\req(tauT)).
\par \noindent 
{\bf Remark}. In the discussion of this subsection, the rapidities $p, p'$ in (\req(pp1c)) satisfy the inequality ${\sf a}^N \neq {\sf b'}^N$, a condition derived from the assumption that the first relation in (\req(tCPM)) is excluded in our consideration above. However, the argument in this subsection about the selfdual solution of (\req(TArel)) using rapidities in (\req(xy0)) or (\req(xy1)) equally holds for $p, p'$ in (\req(pp1c)) satisfying ${\sf a}^N = {\sf b'}^N$ (which implies ${\sf a'}^N = {\sf b}^N$). Since a $\tau^{(2)}$-model satisfying the first relation in (\req(tCPM)) can be reduced to the $\tau^{(2)}$-model with parameters satisfying (\req(ab0)) or (\req(ab2)) by the gauge transform (\req(nu)), one may also use the transfer matrices of the selfdual Potts model to construct the $Q_R$, $Q_L$-operator of  $\tau^{(2)}$-models with the first relation in (\req(tCPM)), in which case we has also previously discussed the $Q_R$, $Q_L$-operator in \cite{R075} through the superintegrable CPM transfer matrices $T_{p, p'}$ and $\widehat{T}_{p, p'}$.

\section{Degenerate chiral Potts models for ${\sf c}^N \neq 1$ \label{sec:dCP}}
\setcounter{equation}{0}
This section is devoted to the study of $Q$-operator of the $\tau^{(2)}$-models with ${\sf c}^N \neq 1$. By Theorem \ref{thm:CPt}, we need only to consider the cases not covered by the second inequality in (\req(tCPM)), i.e. the parameters (\req(Lab')) satisfying the condition (\req(nz)) with ${\sf c}^N \neq 1$ and 
\be
({\sf a'}^N - {\sf b}^N)({\sf a}^N - {\sf b'}^N) ({\sf b}^N {\sf b'}^N- {\sf c}^N{\sf a}^N{\sf a'}^N) = 0.
\ele(sCn1)
The above equality (\req(sCn1)) can be replaced by either ${\sf b}^N {\sf b'}^N- {\sf c}^N{\sf a}^N{\sf a'}^N = 0$, or $({\sf a'}^N - {\sf b}^N)({\sf a}^N - {\sf b'}^N) = 0$ (equivalent to ${\sf b}^N {\sf b'}^N- {\sf c}^N{\sf a}^N{\sf a'}^N  \neq 0$  by (\req(nz))). In subsection \ref{ssec.k'1}, we shall show the $Q$-operator with ${\sf b}^N {\sf b'}^N- {\sf c}^N{\sf a}^N{\sf a'}^N=0$ is given by the transfer matrix of the degenerate chiral Potts models for $k'=1$. When $({\sf a'}^N - {\sf b}^N)({\sf a}^N - {\sf b'}^N) = 0$, we illustrate in subsection \ref{ssec.k'0} that the standard construction of a commuting family of $Q$-operators  in \cite{B72, R075} fails in this case albeit one can obtain the $Q_R$, $Q_L$-operator through the theory of degenerate chiral Potts model for $k'=0$.

\subsection{The degenerate chiral Potts model with $k'=1$  \label{ssec.k'1}}
In this subsection, we construct the $Q$-operator of $\tau^{(2)}$-models with  (\req(nz)), ${\sf c}^N \neq 1$ and ${\sf b}^N {\sf b'}^N- {\sf c}^N{\sf a}^N{\sf a'}^N=0$. By (\req(nz)),  ${\sf a'}^N \neq {\sf b}^N, {\sf a}^N \neq {\sf b'}^N$. Using relations (\req(nu)) and (\req(lambda)), one may assume
\be
{\sf c}^N {\sf a'}^N- {\sf b}^N={\sf c}^N{\sf a}^N -  {\sf b'}^N= {\sf c}^N - 1  \Longleftrightarrow \ {\sf c}^N = \frac{1 - {\sf b'}^N}{1- {\sf a}^N  } = \frac{1 - {\sf b}^N}{1- {\sf a'}^N  }.
\ele(Nab)
By the coordinates $(x , y)$ with $x^N= \xi^N, t= xy$ in (\req(dxi)), we obtain 
\be
x^N + y^N = x^N y^N  ~ \Longleftrightarrow  ~ (1- x^N)^{-1} = 1- y^N =: \mu^N .
\ele(x+y)
The variable $\mu^N$ is related to $t$ by
$$
t^N = (1 - \mu^N ) (1- \mu^{-N}).
$$
The relation (\req(Nab)) implies $\frac{1}{{\sf a}^N} + \frac{1}{{\sf b}^N} = \frac{1}{{\sf a'}^N} + \frac{1}{{\sf b'}^N}$, which is equal to $1$ by using ${\sf c}^N = \frac{{\sf b}^N {\sf b'}^N}{{\sf a}^N{\sf a'}^N}$ and ${\sf a'}^N \neq {\sf b}^N$. Therefore $({\sf a}, {\sf b}), ({\sf a'}, {\sf b'})$ satisfy the first relation in (\req(x+y)). Let $p, p'$ be the elements in (\req(x+y))  defined by
\be
p: (x_p, y_p, \mu_p) = ({\sf a}, {\sf b},  (1-{\sf b}^N)^{\frac{1}{N}}) , \ ~ p': (x_{p'}, y_{p'}, \mu_{p'})= ({\sf a'}, {\sf b'},  {\sf c}(1-{\sf b}^N)^{\frac{-1}{N}}).
\ele(ppNew)
Note that by ${\sf c}^N = (1- {\sf b}^N)(1-{\sf b'}^N)$, the above $\mu_{p'}$ differs from 
$(1-{\sf b'}^N)^{\frac{1}{N}}$ only by a $N$th root of unity. 
The curve (\req(x+y)) can be regarded as the rapidity curve for $k'=1$ in (\req(rapidC)) where the variables $(a, b, c, d)$ is replaced by $(\sqrt[N]{k k'^{-1}} a, \sqrt[N]{k k'^{-1}} b, c, d)$, by which   
the Boltzmann weights (\req(CPW)) with rapidities in (\req(x+y)) give rise to a solution  of (\req(TArel)). Hence the relations 
(\req(xixi*))-(\req(QTpp)) for $q$ in (\req(x+y)) define the transfer matrices $T_{p, p'}(q), \widehat{T}_{p, p'}(q)$ of the degenerate $k'=1$ chiral Potts model, which provide the $Q_R$, $Q_L$-operators of the $\tau^{(2)}$-model satisfying  the commutation relation (\req(TTc)) and 
$\tau^{(2)}T$-relation (\req(tauT)). \par \noindent 
{\bf Remark}. In the above discussion, we assume ${\sf c}^N \neq 1$. However, the described $Q$-operator construction in above is also valid for the case ${\sf c}^N=1$ when $p, p'$ in (\req(ppNew)) are elements in the curve (\req(x+y)) with non-zero ${\sf b}, {\sf b'}$ satisfying $(1- {\sf b}^N)(1-{\sf b'}^N)=1$.

\subsection{The degenerate $\tau^{(2)}$-model for $k'=0$ \label{ssec.k'0}}
We now study the case: ${\sf c}^N \neq 1$ with
$({\sf a'}^N - {\sf b}^N)({\sf a}^N - {\sf b'}^N) = 0$. By relations (\req(nu)) and (\req(lambda)), one may assume one of the following cases holds:
$$
{\sf a'}^N = {\sf b}^N = 1, \ {\sf b'}^N -{\sf c}^N{\sf a}^N  = 1-{\sf c}^N  ~  \ {\rm or} \ ~ ~ \ ~ {\sf a}^N = {\sf b'}^N=1 , {\sf b}^N -{\sf c}^N {\sf a'}^N = 1-{\sf c}^N , 
$$
which imply ${\sf c}^N = \frac{ 1-{\sf b'}^N}{1- {\sf a}^N}$ or $ \frac{ 1-{\sf b}^N}{1- {\sf a'}^N}$ respectively. 
With the coordinates $(x, y)$ with $x^N=\xi^N, t=xy$, the relation (\req(dxi)) becomes 
\be
1 - y^N - x^N +  x^Ny^N = 0 . 
\ele(ck'0)
The above equation can be regarded the $k'=0$ limit of (\req(xyx)) (or (\req(rapidC)))\footnote{By changing $k' \mu^{\pm N}$ by $\mu^{\pm N}$ in (\req(xymu)), the $k'=0$ limit of ${\goth W}_{k'}$ is ${\goth C}_\pm $ respectively.} , which is composed of the two curves
\bea(ll)
{\goth C}_+ : x^N=1, \  \mu^N= 1- y^N , & {\goth C}_-:  y^N=1, \ \mu^{-N} = 1- x^N , 
\elea(C12)
whose element is denoted by $\sigma= (x_\sigma, y_\sigma, \mu_\sigma)$. As in (\req(CPW)), we define the following weights for certain rapidities in (\req(C12)) with $\sigma \in {\goth C}_i, \sigma' \in {\goth C}_j$ for $i, j = \pm $:
\bea(ll)
W_{\sigma, \sigma'}(n) = (\frac{\mu_\sigma}{\mu_{\sigma'}})^n \prod_{j=1}^n
\frac{y_{\sigma'}-\omega^j x_\sigma}{y_\sigma- \omega^j x_{\sigma'} }, & {\rm if} \ i=j , \\ 
\overline{W}_{\sigma, \sigma'}(n)  = ( \mu_\sigma\mu_{\sigma'})^n \prod_{j=1}^n \frac{\omega x_\sigma - \omega^j x_{\sigma'} }{ y_{\sigma'}- \omega^j y_\sigma } & {\rm if} \ i \neq j.
\elea(wk'0)
Note that for elements $\sigma, \sigma'$ in (\req(C12)) with $x$ or $y= 1$ where $\mu$ takes the zero or $\infty$, the above Boltzmann weights are uniquely determined. However the formula in (\req(wk'0)) are not defined when the indices $i, j$ are not in the above described regions, hence the weights in (\req(wk'0)) {\it do not} provide a solution of the star-triangle relation (\req(TArel)). Define the following elements $p, p'$ in (\req(C12)):
$$
\begin{array}{ll} (x_p, y_p, \mu_p) = ({\sf a}, {\sf b}, (1-{\sf a}^N)^\frac{-1}{N}) \in {\goth C}_-,  (x_{p'}, y_{p'}, \mu_{p'}) = ({\sf a'}, {\sf b'}, {\sf c}(1-{\sf a}^N)^\frac{1}{N})  \in {\goth C}_+ & {\rm if} \ {\sf a'}^N = {\sf b}^N = 1, \\
(x_p, y_p, \mu_p) = ({\sf a}, {\sf b}, {\sf c} (1-{\sf a'}^N)^\frac{1}{N}) \in {\goth C}_+ , \ (x_{p'}, y_{p'}, \mu_{p'}) = ({\sf a'}, {\sf b'}, (1-{\sf a'}^N)^\frac{-1}{N}) \in {\goth C}_-  & {\rm if} \ {\sf a}^N = {\sf b'}^N = 1 . 
\end{array} 
$$
One may still use the relations 
(\req(xixi*))-(\req(QTpp)) with $q \in {\goth C}_1 \cup {\goth C}_2 $ to define the $Q_R$, $Q_L$-operator satisfying the relation (\req(tauT)), which are identified with $T_{p, p'}(q), \widehat{T}_{p, p'}(q')$ in (\req(Tpq)), (\req(hTpq)) for $q \in {\goth C}_i, q' \in {\goth C}_j$ and  $i \neq j$. Note that there is no common rapidity variable $q$ valid for both $T_{p, p'}$ and $\widehat{T}_{p, p'}$.  Due to the lack of the star-triangle relation for weights in (\req(wk'0)),  the commutation relation (\req(TTc)) fails in this situation, which prevent us  to obtain the commuting family of $Q$-operators.

In the $Q$-operator discussion of this paper, we assume the condition (\req(nz)). From the general "rapidity" constraint (\req(dxi)), one derives the rapidity $xy$-curves in (\req(xy0)) (\req(xy1)) (\req(x+y)) of the star-triangle solutions (\req(TArel)),
 which are all symmetrical when interchanging $x$ and $y$. Indeed such symmetric property is encoded in the theory since the substitutions (\req(xixi*)) in constructing $Q_R$, $Q_L$-operator, and the commuting $Q$-operators, require all the operators should share the same curve of rapidities. In case the condition (\req(nz)) fails, i.e. ${\sf c}^N \neq 1$ with ${\sf c}^N {\sf a'}^N  = {\sf b}^N$ (or ${\sf c}^N {\sf a}^N  = {\sf b'}^N$), the relation (\req(dxi)), other than the solution $\xi=0$, enables us to derive the curve $ 1 - x^N + x^Ny^N = 0$ through a similar procedure as before. However, the non-symmetric nature of the $xy$-curve prohibits the connection between those $\tau^{(2)}$-models and CPM. Indeed the cyclic-vector construction in (\req(kv)) (\req(ckv)) only leads to the pseudo-vacuum state which serves a simple reference state acted iteratively by the "creation" B-operator in the algebraic Bethe ansatz method to produce a simultaneously diagonalized basis of the $\tau^{(2)}$-operator. By this, one may regard Baxter's $Q$-operator method in CPM and the ABCD-algebra method in algebraic Bethe ansatz are {\it complementary} techniques in the theory of generalized $\tau^{(2)}$-models.

\section{Functional relations of a degenerate chiral Potts model for $k'=1$} \label{sec:FRdCP}
By the discussion in subsection \ref{ssec.sdCP} and subsection \ref{ssec.k'1}, the Boltzmann weights (\req(CPW)) with rapidities in (\req(xy0)), (\req(xy1)) or (\req(x+y)) are solutions of the star-triangle relation (\req(TArel)), which define the degenerate chiral Potts model with $k'=1$.  By the same argument,  the functional relations of CPM in \cite{BBP} indeed also hold for these degenerate models with $k'=1$, which we now explain below. First note that each of the rapidity curves, (\req(xy0)) (\req(xy1)) or (\req(x+y)), is  invariant under automorphisms in (\req(Aut)), and $T_{p, p'}(q), \widehat{T}_{p, p'}(q)$ are single-valued functions of $x_q$ and $y_q$, which will also be denoted by    
$T_{p, p'}(x_q, y_q), \widehat{T}_{p, p'}(x_q, y_q)$ as in the CPM case. By the construction of the $Q$-operator, the $\tau^{(2)}T$-relation (\req(tauT)) holds for those degenerated models. Indeed the arguments  in deriving functional equations of CPM,  and formulas (3.13)-(4.45) in \cite{BBP}  are all valid for these degenerate chiral Potts models for $k'=1$. The fusion matrix $\tau^{(j)} (t_q)$ in subsection \ref{ssec.tau2} of this paper is the same as $\tau^{(j)}_{k, q}$ in \cite{BBP} (3.44a) with $k=0$ : $\tau^{(j)}_{0, q} = \tau^{(j)} (t_q)$; the fusion relations, (\req(fus)) and (\req(bfu)), are given by formulas $(4.27a)_{k=m=0}$,  $(4.27c)_{k=0}$ respectively, in \cite{BBP}. The $\tau^{(2)}T$-relation (\req(tauT)) is the same as \cite{BBP}  (4.20) (4.21) (for $k=0$ and setting (2.41) $\xi_q = \hat{\xi}_q =1$). The $\tau^{(j)}T$-relation is now expressed by (\cite{BBP} $(4.34)_{k=0}$)
$$
\begin{array}{l}
\tau^{(j)}(t_q)= \sum_{m=0}^{j-1} \varphi_q \varphi_{Uq} \cdots \varphi_{U^{m-1}q}
\overline{\varphi}_{U^{m+1}q} \overline{\varphi}_{U^{m+2}q}  \cdots \overline{\varphi}_{U^{j-1}q} \\
T_{p, p'}(x_q, y_q)T_{p, p'}(\omega^m x_q, y_q)^{-1} T_{p, p'}(\omega^j x_q,y_q) T_{p, p'}(\omega^{m+1} x_q, y_q)^{-1}X^{j-m-1}
\end{array}
$$
where $\varphi_{U^{-1}q}, \overline{\varphi}_q$ are the scale-factors in the $\tau^{(2)}T$-relation (\req(tauT')): 
$\varphi_q = \{\frac{(y_p-  \omega x_q)(t_{p'}- t_q) }{y_p y_{p'}(x_{p'}-  x_q)}\}^L $, $\
\overline{\varphi}_q = \{\frac{\omega \mu_{p'} \mu_p(t_p- t_q)(x_{p'}- x_q) }{y_p y_{p'}(y_p- \omega x_q)}\}^L $, which are related to $z(t)$ in (\req(fus)) by $z(t_q)= \varphi_q \overline{\varphi}_q$. The relation of CPM transfer matrix and $\tau^{(j)}$'s is given by the $T\hat{T}$-relation ( \cite{BBP}  $(3.46)_{k=0}$ ):
$$
\begin{array}{l}
\lambda_q^{(0, j)} T_{p, p'}(x_q, y_q) \widehat{T}_{p, p'}(y_q, \omega^j x_q) = \overline{H}_{p' q}^{(j)} \tau^{(j)} (t_q) + H_{p q}^{(j)} \tau^{(N-j)} (\omega^j t_q) X^j 
\end{array}
$$
where  $\overline{H}_{p' q}^{(j)}  = (\frac{\omega^{j(j-1)/2} t_{p'}^{j-N} \prod_{l=j}^{N-1}(t_{p'}- \omega^l t_q)}{(1- x_q^N/x_{p'}^N)(- x_{p'}\mu_{p'} \mu_q/ y_{p'})^j} )^L$, $H_{p q}^{(j)}= (\frac{\omega^{j(j+1)/2} y_p^{-j} \prod_{l=0}^{j-1}(t_p- \omega^l t_q)}{(1- x_q^N/y_p^N)(- y_p \mu_q/ \mu_p)^j} )^L $, $\lambda_q^{(0, j)} = (N\Omega_{pq}^{0j}\overline{\Omega}_{p' q}^{0j})^{-L}$ with $\Omega_{pq}^{0j}= \frac{y_p^{j-1}}{\prod_{l=1}^{j-1} (y_p - \omega^l x_q) }$ and $\overline{\Omega}_{p' q}^{0j}= \frac{(\mu_{p'} \mu_q)^j y_{p'}^{N-j-1}(y_{p'} - y_q)\prod_{l=0}^{j-1} (x_q - \omega^{-l} x_{p'})}{(y_{p'}^N - y_q^N ) }$  (\cite{BBP} (3.24) (3.35) (3.36) (3.41) (3.42)). Using formulas (4.37)-(4.38) in \cite{BBP}, one can write $T\hat{T}$-relation  in the form (\cite{B02} (13), \cite{R05o} (15)):
$$
\begin{array}{l}
 T_{p, p'}(x_q, y_q) \widehat{T}_{p, p'}(y_q, \omega^j x_q) = r_{p', q} h_{j; p, p',q} \bigg(\tau^{(j)} (t_q) + \frac{z(t_q)z(\omega t_q) \cdots z(\omega^{j-1} t_q)}{\alpha_q } \tau^{(N-j)} (\omega^j t_q) X^j \bigg) 
\end{array}
$$
where  $r_{p', q}= (\frac{N(x_{p'}-x_q)(y_{p'}-y_q)(t_{p'}^N -t_q^N)}{(x_{p'}^N- x_q^N) (y_{p'}^N -y_q^N)(t_{p'}-t_q)})^L $, $h_{j; p, p', q}=( \prod_{m=1}^{j-1} \frac{y_p y_{p'} (x_{p'}- \omega^m x_q)}{(y_p-\omega^m x_q)(t_{p'}-\omega^m t_q)} )^L $, and $\alpha_q$ is in (\req(eeq)).
In particular for $j=N$, the $T\hat{T}$-relation reduces to (\cite{BBP} (4.44) )
$$
T_{p, p'}(x_q, y_q) \widehat{T}_{p, p'}(y_q, x_q) = \bigg( \frac{N(y_p y_{p'})^{N-1}(y_p-x_q)(y_{p'}-y_q)}{ (y_p^N - x_q^N) (y_{p'}^N - y_q^N)} \bigg)^L  \tau^{(N)}(t_q) 
$$
Then  one can derive the functional relation of CPM transfer matrix (\cite{BBP}(4.40)):
$$
\widehat{T}_{p, p'}(y_q, x_q) = \sum_{m=0}^{N-1} C_{m, q} T_{p, p'}(\omega^m x_q, y_q)^{-1}T_{p, p'}(x_q, y_q) T_{p, p'}(\omega^{m+1} x_q, y_q)^{-1} X^{-m-1}
$$
where $C_{m, q} = \varphi_q \varphi_{Uq} \cdots \varphi_{U^{m-1}q}
\overline{\varphi}_{U^{m+1}q} \overline{\varphi}_{U^{m+2}q}  \cdots \overline{\varphi}_{U^{N-1}q}( \frac{N(y_p y_{p'})^{N-1}(y_p-x_q)(y_{p'}-y_q)}{ (y_p^N - x_q^N) (y_{p'}^N - y_q^N)} )^L $.

\section{Concluding Remarks}\label{sec.F} 
Through the $Q$-operator approach, we establish the equivalent relation between the theories of generalized $\tau^{(2)}$-model and the $N$-state chiral Potts models with the degenerate forms included. The application of a special gauge transform and the rescaling of spectral parameters of the $L$-operator has effectively deduced the five-parameter $\tau^{(2)}$-family to the three-parameter ones in CPM. The "generic" $\tau^{(2)}$-models correspond to CPM with two vertical rapidities in ${\goth W}_{k'}$ with $ k' \neq 0, \pm 1$ in (\req(rapidC)), and the result is verified by an algebraic-geometry method. The explicit form of the generic parameters is described in Theorem \ref{thm:CPt}. Other than a special kind of $\tau^{(2)}$-models  (\req(aba)) which can be treated by the algebraic Bethe ansatz method, the Baxter's $Q_{72}$-operator technique is successfully applied to the rest "non-generic"  $\tau^{(2)}$-models, where the $Q_R$, $Q_L$-operators are represented by transfer matrices $T_{p, p'}, \widehat{T}_{p, p'}$ of the degenerate chiral Potts model for $k' = 1, 0$. The degenerate models for $k'=1$ all arise from the selfdual solutions of the star-triangle relation (\req(TArel))  \cite{AMPT, MPTS, BPA, FatZ}. 
As a result of our working, an explicit matrix form of the $\tau^{(2)}$-model is found, and functional relations are verified for the selfdual Potts models in the same way as the solvable CPM in \cite{BBP}.  It would be desirable that the functional-relation method can also be employed in the investigation of eigenvalue problem for those degenerate models, just  as in the discussion of CPM in \cite{B90, B93, MR}. A programme along this line is now under progress and partial results are promising.

\section*{Acknowledgements} 
The author is pleased to thank Laboratoire de Mathematiques et Physique Theorique, CNRS/UMR, University of Tours, France 
for the hospitality in the fall of 2007, where part of this work was carried out. He also wishes to acknowledge many fruitful discussions with Professor P. Baseilhac.
This work is supported in part by National Science Council of Taiwan under Grant No NSC 96-2115-M-001-004.

\section*{Appendix: Algebraic geometry of chiral Potts $\tau^{(2)}$-models with two alternating rapidities}
\setcounter{equation}{0}
In this appendix, we provide an algebraic geometry proof of Theorem \ref{thm:CPt}. First we determine the explicit equations about parameters  in (\req(Lab')) corresponding to the 3-parameter CPM family  (\req(tpp')). For simple notations, in this appendix we shall write $x=x_p, y= y_p, \mu= \mu_p, x'=x_{p'}, y'=y_{p'},  \mu'= \mu_{p'}$ for $p, p' \in {\goth W}_{k'}$, then the relation (\req(xyx)) yields 
\be
k = \frac{x^N + y^N}{1 + x^N y^N } = \frac{x'^N + y'^N}{1 + x'^N y'^N}, \ \ \ \  \mu^{-N}   = \frac{1 -  k x^N}{k'} , \ \ \ \mu'^{-N}   = \frac{1 -  k x'^N}{k'} \tag{A1}
\ele(xyx'y')
with the condition about $k \neq 0, \pm 1, \infty $, which is equivalent to the constraints, 
\be
\left\{ \begin{array}{ll} (x^N + y^N)(1-x^{2N})(1-y^{2N})(1+x^Ny^N) \neq 0, & {\rm or} \ ~ \ x^N = - y^N = \pm 1 , \\
 (x'^N + y'^N)(1-x'^{2N})(1-y'^{2N})(1+x'^Ny'^N) \neq 0, & {\rm or} \ ~ \ x'^N = - y'^N = \pm 1 . \\
\end{array} \right. \tag{A2}
\ele(CPc)
The CPM condition (\req(tpp')) now becomes
\be
({\sf a, b, a', b', c}) = (x, y, x', y', \mu \mu'). \tag{A3}
\ele(txx')
By  (\req(xyx'y')), one finds 
$$
y'^N = \frac{(x^N + y^N) - (1+ x^Ny^N) x'^N}{(1+ x^Ny^N) - (x^N + y^N)x'^N} , \ ~ \
(\mu \mu')^{-N}   
= \frac{1 + x^N y^N -  (x^N + y^N)x'^N }{1  -  y^{2N}}, 
$$
equivalently, the elements in (\req(txx')) satisfy the relations
\bea(l)
(1+ {\sf a}^N {\sf b}^N){\sf b'}^N  - ({\sf a}^N + {\sf b}^N){\sf a'}^N{\sf b'}^N = ({\sf a}^N + {\sf b}^N) - (1+ {\sf a}^N {\sf b}^N) {\sf a'}^N , \\
(1 + {\sf a}^N {\sf b}^N){\sf c}^N    -  ({\sf a}^N + {\sf b}^N){\sf a'}^N{\sf c}^N   
= 1  -  {\sf b}^{2N}, \tag{A4}
\elea(CPeq)
with the constraint condition (\req(CPc)) replaced by 
\bea(ll)
{\rm either }& (x^N + y^N)(1+x^Ny^N)(1-x^{2N})(1-y^{2N})(1-x'^{2N}) \neq 0, \\
{\rm or} & x^N = - y^N = \pm 1 , \\
{\rm or} & (x^N + y^N)(1-x^{2N})(1-y^{2N})(1+x^Ny^N) \neq 0 \ ,  x'^N= \pm 1. \tag{A5}
\elea(CPc2)
The first case in above implies $(x'^N + y'^N)(1+x'^Ny'^N)(1-x'^{2N})(1-y'^{2N}) \neq 0$. For the second case in (\req(CPc2)), one has $\mu^N= (\frac{1 \pm k}{1 \mp  k })^{1/2}$,  which by (\req(xyx'y')), yields $
k  = \frac{1- (\mu \mu')^{-N}}{ x'^N \pm  (\mu \mu')^{-N} }  =  \frac{1- (\mu \mu')^N}{y'^N  \mp (\mu \mu')^N }$.
Then either  $(\mu \mu')^N = 1$ where $- x'^N =y'^N= \pm 1 $,  or $(\mu \mu')^N \neq 1$  where  $x'^N \neq \mp (\mu \mu')^{-N}, \mp 1,  y^N \neq \pm (\mu \mu')^N, \pm 1$ and 
$(\mu \mu')^N  = \frac{1  \pm  y'^N}{1 \mp x'^N}$. For the third case in (\req(CPc2)), one has 
$x'^N= \pm 1$, hence $y'^N = \mp 1 $, and $(\mu \mu')^N   = \frac{(1  \pm  y^{N})  }{(1 \mp  x^N ) }$ by (\req(CPeq)). Therefore equation (\req(CPeq)) subject to the constraint (\req(CPc2)) can be divided into the following cases:
\bea(ll)
(i) & {\sf c}^N = 1 , ~ ~ {\sf a}^N = - {\sf b}^N = - {\sf a'}^N =  {\sf b'}^N= \pm 1 ;  \\
 (ii) & ({\sf a}^N + {\sf b}^N)(1+{\sf a}^N{\sf b}^N)(1-{\sf a}^{2N})(1-{\sf b}^{2N})(1-{\sf a'}^{2N}) \neq 0, \\
& \left\{ \begin{array}{ll}  {\sf a}^N + {\sf b}^N + ({\sf a}^N + {\sf b}^N){\sf a'}^N {\sf b'}^N= {\sf a'}^N+ {\sf b'}^N + ({\sf a'}^N + {\sf b'}^N  ){\sf a}^N {\sf b}^N  ,\\
(1 + {\sf a}^N {\sf b}^N){\sf c}^N    -  ({\sf a}^N + {\sf b}^N){\sf a'}^N{\sf c}^N   
= 1  -  {\sf b}^{2N}; \end{array} \right.
\\
(iii) & {\sf c}^N \neq 1, ~ ~ ~ {\sf a}^N = - {\sf b}^N =  \pm 1,   \ ~ \
({\sf a'}^N  \pm {\sf c}^{-N})({\sf a'}^N  \pm  1) ({\sf b'}^N \mp  {\sf c}^N )({\sf b'}^N \mp  1 )\neq 0, \\
&1 \pm {\sf b'}^N - {\sf c}^N (1 \mp {\sf a'}^N  )= 0 ; \\
(iii') & {\sf c}^N \neq 1, ~ ~ ~ {\sf a'}^N = - {\sf b'}^N =  \pm 1, \ ~ \  ({\sf a}^N \pm {\sf c}^{-N})({\sf a}^N \pm 1)({\sf b}^N \mp {\sf c}^N)({\sf b}^N \mp 1) \neq 0  ,  \\
& 1 \pm {\sf b}^N - {\sf c}^N (1 \mp {\sf a}^N  )= 0. \tag{A6}\\
\elea(2Rap)
The above $(iii)$ and $(iii')$ are symmetrical under the substitution: ${\sf a}, {\sf b} \leftrightarrow  {\sf a'}, {\sf b'}$. The third condition in $(iii)$ is equivalent to  $({\sf a'}^N + {\sf b'}^N)(1-{\sf a'}^{2N})(1-{\sf b'}^{2N})(1+{\sf a'}^N{\sf b'}^N) \neq 0$; a similar statement also exists for $(iii')$. Note that $(ii)$ implies $ ({\sf a'}^N + {\sf b'}^N)(1+{\sf a'}^N{\sf b'}^N)(1-{\sf a'}^{2N})(1-{\sf b'}^{2N})(1-{\sf a}^{2N}) \neq 0$; and when interchanging   
${\sf a} , {\sf b}$ respectively with  $ {\sf a'}, {\sf b'}$ in conditions of $(ii)$, one obtains the equivalent condition for $(ii)$. 

We are going to describe $\CZ^{*2}$-orbits of elements in (\req(2Rap)) for parameters in (\req(Lab')) under the $\CZ^{*2}$-action induced by relations (\req(nu)) and (\req(lambda)), i.e.,  
\be
({\sf a, b, a', b', c}) \mapsto (\lambda \nu^{-1} {\sf a}, \nu {\sf b}, \nu {\sf a}', \lambda \nu^{-1} {\sf b}', {\sf c} ), \ \  \lambda , \nu  \in \CZ^*. \tag{A7}
\ele(nulam)
Denote ${\tt u}:= \lambda^N \nu^{-N}, {\tt v}:=\nu^N \in \CZ^* $. Theorem \ref{thm:CPt} will follow by resolving   
$({\sf a}, {\sf b}, {\sf a'}, {\sf b'} , {\sf c} ) \in \CZ^{*5}$ for each of the following equations (corresponding to those in (\req(2Rap))), so that one can obtain a solution of $({\tt u}, {\tt v}) \in \CZ^{*2}$: 
\bea(ll)
({\rm I}) & {\sf c}^N = 1 , ~ ~ {\tt u} {\sf a}^N = - {\tt v} {\sf b}^N = - {\tt v} {\sf a'}^N =  {\tt u} {\sf b'}^N= \pm 1 ; \\
 ({\rm II}) & ({\tt u} {\sf a}^N + {\tt v} {\sf b}^N)(1+ {\tt u}{\tt v} {\sf a}^N{\sf b}^N)(1-{\tt u}^2 {\sf a}^{2N})(1-{\tt v}^2 {\sf b}^{2N})(1- {\tt v}^2 {\sf a'}^{2N}) \neq 0, \\
&  \left\{ \begin{array}{ll} {\tt u}^2{\tt v} {\sf a}^N {\sf b'}^N({\sf b}^N -  {\sf a'}^N) + {\tt u} {\tt v}^2{\sf b}^N {\sf a'}^N ({\sf a}^N-{\sf b'}^N) = {\tt u}({\sf a}^N - {\sf b'}^N) + {\tt v}({\sf b}^N- {\sf a'}^N ) , \\
{\tt u}{\tt v}{\sf a}^N({\sf b}^N -  {\sf a'}^N){\sf c}^N + {\tt v}^2{\sf b}^N ( {\sf b}^N - {\sf a'}^N {\sf c}^N) = 1 -{\sf c}^N ; \end{array} \right.  \\

({\rm III}) & {\sf c}^N \neq 1, ~ ~ {\tt u} {\sf a}^N = - {\tt v} {\sf b}^N =  \pm 1, ~  ({\tt v} {\sf a'}^N \pm {\sf c}^{-N})({\tt v} {\sf a'}^N \pm 1)
({\tt u}{\sf b'}^N \mp  {\sf c}^N)({\tt u}{\sf b'}^N \mp  1) \neq 0  , \\
& 1 \pm {\tt u} {\sf b'}^N - {\sf c}^N (1 \mp  {\tt v} {\sf a'}^N  )= 0 ; \\

({\rm III}') & {\sf c}^N \neq 1, \ {\tt v}{\sf a'}^N = - {\tt u} {\sf b'}^N =  \pm 1, ~ ~ 
({\tt u} {\sf a}^N  \pm {\sf c}^{-N} ) ({\tt u} {\sf a}^N  \pm 1)({\tt v} {\sf b}^N \mp {\sf c}^N ) ({\tt v} {\sf b}^N \mp 1 ) \neq 0 ,   \\
&1 \pm {\tt v} {\sf b}^N - {\sf c}^N (1 \mp {\tt u}{\sf a}^N  )= 0. \tag{A8}
\elea(2Raps)
Note that the interchange of ${\sf a}, {\sf b}, {\tt u}$ respectively with ${\sf a'}, {\sf b'}, {\tt v}$ leaves the cases  (I) and $({\rm II})$ invariant (only replaced by some equivalent relations), while $({\rm III})$ and $({\rm III}')$ are exchanged.   
The third condition in (III) is equivalent to $({\tt v}{\sf a'}^N + {\tt u}{\sf b'}^N)(1-{\tt v}^2{\sf a'}^{2N})(1-{\tt u}^2{\sf b'}^{2N})(1+{\tt u}{\tt v}{\sf a'}^N{\sf b'}^N) \neq 0$;  a symmetrical statement holds also for $({\rm III}')$. 

When ${\sf c}^N =1$, we need only to consider the cases $({\rm I})$ and $({\rm II})$ in (\req(2Raps)). For the case $({\rm II})$, the second equality equation implies $({\tt u}{\sf a}^N + {\tt v}{\sf b}^N) ( {\sf b}^N - {\sf a'}^N ) = 0$, hence by the first (constraint) condition, ${\sf b}^N = {\sf a'}^N $. Then the first equality equation yields $
 (1- {\tt v}^2{\sf b}^{2N}) ({\sf a}^N-{\sf b'}^N) = 0 $, hence ${\sf a}^N={\sf b'}^N$. Therefore both the cases, $({\rm I})$ and $({\rm II})$ with ${\sf c}^N =1$, satisfy the condition in Theorem \ref{thm:CPt} for ${\sf c}^N = 1$. From now on, we shall assume ${\sf c}^N \neq 1$, where only the cases  $({\rm II})$,  $({\rm III})$ and $({\rm III}')$ to be considered.
First we show 
\be
({\sf b}^N -{\sf a}'^N)({\sf b}'^N -{\sf a}^N )({\sf b}^N -{\sf a'}^N {\sf c}^N) ({\sf b'}^N -{\sf a}^N {\sf c}^N) \neq 0, \tag{A9}
\ele(4nt0)
which is obviously valid for $({\rm III})$ and $({\rm III}')$. Indeed in the case $({\rm II})$, the second equation when ${\sf b}^N ={\sf a'}^N $ or ${\sf b}^N ={\sf a'}^N {\sf c}^N$ implies  $1-{\tt v}^2{\sf b}^{2N}=0$  or $1+{\tt u}{\tt v}{\sf a}^N{\sf b}^N= 0$ respectively, both contradicting the constraint condition. By the symmetrical argument, ${\sf b}'^N \neq {\sf a}^N $ and ${\sf b'}^N \neq {\sf a}^N {\sf c}^N$; hence follows (\req(4nt0)).

Now assume the condition (\req(4nt0)). We are going to study the complex solution $({\tt u}, {\tt v})$ of equations in (\req(2Raps)):
\be 
 \left\{ \begin{array}{ll} {\tt u}^2{\tt v}{\sf a}^N {\sf b'}^N({\sf b}^N -  {\sf a'}^N) + {\tt u} {\tt v}^2{\sf b}^N {\sf a'}^N ({\sf a}^N-{\sf b'}^N) = {\tt u}({\sf a}^N - {\sf b'}^N) + {\tt v}({\sf b}^N- {\sf a'}^N ) , \\
{\tt u}{\tt v}{\sf a}^N({\sf b}^N -  {\sf a'}^N){\sf c}^N + {\tt v}^2{\sf b}^N ( {\sf b}^N -{\sf a'}^N {\sf c}^N) 
= 1 -{\sf c}^N , \end{array} \right. \tag{A10}
\ele(uvsol)
and examine the condition so that the constraints in (\req(2Raps)) are satisfied.  
Note that by ${\sf c} \neq 1$ and ${\sf b}^N \neq {\sf a'}^N $,  any solution of (\req(uvsol)) must have the non-zero ${\tt u}, {\tt v}$-value. Furthermore, the ${\tt u}, {\tt v}$ determined by  $({\tt u}{\sf a}^N,  {\tt v}{\sf b}^N) = \pm ( 1, -1)$  are solutions of (\req(uvsol)), but fail to satisfy the inequality constraint in $({\rm II})$. 
\begin{lem} \label{lem:Cont}
Let $({\tt u}, {\tt v})$ be a solution of $(\req(uvsol))$.  Then the following conditions are equivalent:
\be
({\tt u}{\sf a}^N,  {\tt v}{\sf b}^N) = \pm ( 1, -1) \Longleftrightarrow {\tt v}{\sf b}^N = \mp 1 \Longleftrightarrow {\tt u} {\sf a}^N + {\tt v} {\sf b}^N = 0 \Longleftrightarrow 1+ {\tt u}{\tt v} {\sf a}^N{\sf b}^N = 0 .  \tag{A11}
\ele(Cont)
\end{lem}
{\it Proof}. The first equivalence relation follows from the second equation in (\req(uvsol)) and the condition ${\sf a'}^N \neq {\sf b}^N$. If ${\tt u} {\sf a}^N + {\tt v} {\sf b}^N = 0$, the second equation in (\req(uvsol)) becomes $(1 -{\sf c}^N)(1+{\tt u}{\tt v}{\sf a}^N{\sf b}^N )= 0 $, hence $1+ {\tt u}{\tt v} {\sf a}^N{\sf b}^N=0$. Conversely when $1+ {\tt u}{\tt v} {\sf a}^N{\sf b}^N=0$, one can write the second equation in  (\req(uvsol)) as $({\tt v}^2{\sf b}^{2N}-1) ( {\sf b}^N -{\sf a'}^N {\sf c}^N) = 0$, hence by the assumption ${\sf b}^N \neq {\sf a'}^N {\sf c}^N$,  ${\tt v}{\sf b}^N = \mp 1$. Then follow the results.
\par \noindent 
{\bf Remark}. One can express ${\tt u}{\sf b'}^N$ in terms of ${\tt u}{\sf a}^N, {\tt v}{\sf b}^N , {\tt v}{\sf a'}^N$ using the first equation in (\req(uvsol)), then obtain 
$$
\begin{array}{l}
1 - {\tt u}^2{\sf b'}^{2N} = (1-{\tt u}^2{\sf a}^{2N})(1-{\tt v}^2{\sf a'}^{2N})(1-{\tt v}^2{\sf b}^{2N}) , \\ 
1 + {\sf u}{\tt v}{\sf a'}^N {\sf b'}^N -  ({\tt v}{\sf a'}^N + {\tt u}{\sf b'}^N){\tt u}{\sf a}^N = 
(1-{\tt u}^2{\sf a}^{2N})(1-{\tt v}^2{\sf a'}^{2N})(1 + {\tt u}{\sf v}{\sf a}^N {\sf b}^N -  ({\tt u}{\sf a}^N + {\tt v}{\sf b}^N){\tt v}{\sf a'}^N ),
\end{array} 
$$
which in turn yield the equations symmetrical to those in (\req(uvsol)) by interchanging ${\sf a}, {\sf b}, {\tt u}$ respectively with ${\sf a'}, {\sf b'}, {\tt v}$. Therefore follows the equivalence of (\req(Cont)) and its dual relation, which is obtained by replacing ${\tt u}{\sf a}^N,  {\tt v}{\sf b}^N$ in (\req(Cont)) by ${\tt v}{\sf a'}^N,  {\tt u}{\sf b'}^N$ respectively.
\par  \vspace{.1in}
We now determine the solutions of (\req(uvsol)) other than those in the above lemma. 
By  ${\sf b}^N \neq  {\sf a'}^N$, and the second relation of (\req(uvsol)), we express ${\tt u}$ in terms of ${\tt v}$, substituted in the first equation of (\req(uvsol)), which is now equivalent to the ${\tt v}$-polynomial:  
\be 
C_4 ({\sf b}^N {\tt v})^4 + C_2 ({\sf b}^N{\tt v})^2 + C_0 = 0 , \tag{A12}
\ele(C)
where $C_0 = (1 -{\sf c}^N)({\sf b'}^N - {\sf a}^N {\sf c}^N ) ~ (\neq 0)$, and 
$$
\begin{array}{l}
C_2 =  -2 {\sf b'}^N  + {\sf b}^{-N} ({\sf a}^N+ {\sf b'}^N )  ({\sf b}^N+ {\sf a'}^N   ) {\sf c}^N    - {\sf a}^N  {\sf b}^{-2N}({\sf b}^{2N}+ {\sf a'}^{2N})  {\sf c}^{2N} , \\
C_4 = {\sf b}^{-2N}( {\sf b}^N -{\sf a'}^N {\sf c}^N) ( {\sf b}^N {\sf b'}^N -  {\sf a}^N {\sf a'}^N {\sf c}^N  ) .
\end{array}
$$
Note that $C_0+C_2+C_4=0$. Claim: $ {\sf b}^N {\sf b'}^N \neq  {\sf a}^N {\sf a'}^N {\sf c}^N$. Otherwise, $C_4=0$, which implies  ${\tt v}{\sf b}^N = \mp 1$, hence ${\tt u}{\sf a}^N= \pm 1$ and ${\tt u}{\sf b'} + {\tt v}{\sf a'}{\sf c}=0$, contradicting to the conditions in $({\rm II})$,$({\rm III})$ and $({\rm III}')$. Therefore (\req(C)) is a fourth-order equation with the solutions given by 
\be
{\tt v}^2 {\sf b}^{2N} = 1 ~ {\rm or} ~ ~ \frac{{\sf b}^{2N} (1 - {\sf c}^N)( {\sf b'}^N - {\sf c}^N {\sf a}^N )}{( {\sf b}^N - {\sf a'}^N {\sf c}^N) ( {\sf b}^N  {\sf b'}^N  -  {\sf a}^N {\sf a'}^N {\sf c}^N   ) }. \tag{A13}
\ele(solC)
Using ${\sf a'}^N \neq {\sf b}^N$, the condition of the above second solution with value $1$ is equivalent to the relation
${\sf a}^N  {\sf b}^N (1-{\sf c}^N) +   {\sf b}^N  {\sf b'}^N -  {\sf a}^N {\sf a'}^N {\sf c}^N =   0$. By Lemma \ref{lem:Cont}, $({\tt u}{\sf a}^N,  {\tt v}{\sf b}^N) = \pm ( 1, -1)$. Then follows the equality relation in $({\rm III})$, where the constraint inequalities also hold by the remark of Lemma \ref{lem:Cont}. 
We now consider the case when ${\tt v}^2 {\sf b}^{2N}$ is given by the second solution in (\req(solC)) which is not equal to one. By Lemma \ref{lem:Cont}, 
${\tt v} {\sf b}^{N}$ gives rise a solution of the case $({\rm II})$ in (\req(2Raps)) except the constraint condition ${\tt v}^2 {\sf a'}^{2N} \neq 1$. 
In case ${\tt v} {\sf a'}^N = \pm 1$, the second equation of (\req(uvsol)) becomes ${\sf c}^N(1 \mp {\tt u}{\sf a}^N  )(1 \mp  {\tt v}{\sf b}^N ) = 1- {\tt v}^2{\sf b}^{2N}$, and ${\tt u} {\sf b'}^N = \mp 1$ holds by the remark of Lemma \ref{lem:Cont}. Since ${\tt v}^2 {\sf b}^{2N} \neq 1$, this provides a solution of $({\rm III})$ in (\req(2Raps)). This completes the proof of Theorem \ref{thm:CPt}.


\begin{thebibliography}{99}
\bibitem{AMP} G. Albertini, B. M. McCoy, and 
J. H. H. Perk, Eigenvalue spectrum of the
superintegrable chiral Potts model, in  Adv. Stud.
Pure Math., 19, Kinokuniya Academic (1989) 1--55.
%
\bibitem{AMPTY} H. Au-Yang, B. M. McCoy,  
J. H. H. Perk, S. Tang and M. L. Yan, Commuting transfer matrices in chiral Potts models: solutions of the star-triangle equations with genus $> 1$, Phys. Lett. A 123 (1987) 219--223.
%
\bibitem{AMPT} H. Au-Yang, B. M. McCoy,  
J. H. H. Perk and S. Tang, Solvable models in statistical mechanics and Riemann surfaces of genus greater than one, {\it Algebraic Analysis}, Vol. 1 , eds. M. Kashiwara and T. Kawai, Academic Press, San Diego (1988), 29--40.
%
\bibitem{B72} R. J. Baxter, Partition function of the eight vertex model, Ann. Phys. 70 (1972) 193--228.
%
\bibitem{B73} R. J. Baxter, Eight-vertex model in lattice statistic and  one-dimensional anisotropic Heisenberg chain I: Some fundamental eigenvalues,  II. Equivalence to a generalized Ice-type lattice model,  III. Eigenvalues of the transfer matrix and Hamiltonian, Ann. Phys. 76 (1973) 1--71. 
%
\bibitem{Bax} R. J. Baxter, Exactly solved models
in statistical mechanics, Academic Press (1982).
%
\bibitem{B89} R. J. Baxter, Superintegrable chiral Potts model: Thermodynamic properties, an "Inverse" model, and a simple associated Hamiltonian, J. Stat. Phys. 57 (1989) 1--39.
%
\bibitem{B90} R. J. Baxter, Chiral Potts model: eigenvalues of the transfer matrix, Phys. Lett. A 146 (1990) 110--114.
%
\bibitem{B93} R. J. Baxter, Chiral Potts model with skewed boundary conditions, J.
Stat. Phys. 73 (1993) 461--495.
%
\bibitem{B02} R. J. Baxter, The "inversion relation" method for obtaining the free energy of the chiral Potts model, Physica A 322 (2003) 407--431; cond-mat/02121075.
%
\bibitem{B049} R. J. Baxter, Transfer matrix functional relation for the generalized $\tau_2(t_q)$ model, J. Stat. Phys. 117 (2004) 1--25;  cond-mat/0409493.
%
\bibitem{BBP} R. J. Baxter, V.V. Bazhanov and
J.H.H. Perk,  Functional relations for transfer
matrices of the chiral Potts model, Int. J. Mod.
Phys. B 4 (1990) 803--870.
%
\bibitem{BPA} R. J. Baxter, J. H. H. Perk and H. Au-Yang, New solutions of the  star-triangle relations for the chiral Potts model, Phys. Lett. A 128 (1988) 138--142.
%
\bibitem{BazS} V.V. Bazhanov and Yu.G. Stroganov, Chiral
Potts model as a descendant of the six-vertex model, J.
Stat. Phys. 59 (1990) 799--817.
%
\bibitem{DFM} T. Deguchi, K. Fabricius and B. M. McCoy, The $sl_2$ loop algebra symmetry for the six-vertex model at roots of unity, J. Stat. Phys. 102 (2001) 701--736; cond-mat/9912141. 
%
\bibitem{FM01} K. Fabricius and B. M. McCoy, Evaluation parameters and Bethe roots for the six vertex model at roots of unity, {\it Progress in Mathematical Physics} Vol 23, eds. M. Kashiwara and T. Miwa,  Birkh\"{a}user Boston (2002), 119--144; cond-mat/0108057.
%
\bibitem{FM02} K. Fabricius and B. M. McCoy, New developments in the eight vertex model, J. Stat. Phys. 111 (2003) 323--337; cond-mat/0207177.
%
\bibitem{FM04} K. Fabricius and B. M. McCoy, Functional equations and fusion matrices for the eight vertex model, Publ. RIMS, 40 (2004) 905--932; cond-mat/0311122.
%
\bibitem{F06} K. Fabricius, A new $Q$-operator in the eight-vertex model, J. Phys. A: Math. Theor. 40 (2007) 4075--4086; cond-mat/0610481v3.
%
\bibitem{Fad} L. D. Faddeev, How algebraic Bethe
Ansatz works for integrable models, eds. A.
Connes, K. Gawedzki and J. Zinn-Justin, {\it
Quantum symmetries/ Symmetries quantiques},
Proceedings of the Les Houches summer school,
Session LXIV, Les Houches, France, August 1-
September 8, 1995, North-Holland (1998),  149--219;
%
\bibitem{FatZ} V. A. Fateev and A. B. Zamolodchikov, Self-dual solutions of the star-triangle relations in $\ZZ_N$-models, Phys. Lett. A 92 (1982) 37--39.
%
\bibitem{GIPS} G. von Gehlen, N. Iorgov, S. Pakuliak and V. Shadura: Baxter-Bazhanov- Stroganov model: Separation of variables and Baxter equation, J. Phys. A: Math. Gen. 39 (2006) 7257--7282; nlin.SI/0603028.
%
\bibitem{KBI} V. E. Korepin, N. M. Bogoliubov, and A. G. Izegin, Quantum inverse scattering method and correlation functions, Cambridge Univ. Press, Cambridge, 1993.
%
\bibitem{KS} P. P. Kulish and E. K. Sklyanin,
Quantum spectral transform method. Recent
developments, eds. J. Hietarinta and C. Montonen,
Lecture Notes in Physics 151 Springer (1982),
61--119.
%
\bibitem{MPTS} B. M. McCoy,  
J. H. H. Perk, S. Tang and C. H. Sah, Commuting transfer matrices for the four-state self-dual chiral Potts model with a genus-three uniformizing Fermat curve, Phys. Lett. A 125 (1987) 9--14.
%
\bibitem{MR} B. M. McCoy and S. S. Roan, Excitation spectrum and phase structure of the chiral Potts model. Phys. Lett. A 150 (1990) 347--354.
%
\bibitem{R04} S. S. Roan, Chiral Potts rapidity curve descended from six-vertex model and symmetry group of rapidities, J. Phys. A: Math. Gen. 38 (2005) 7483--7499; cond-mat/0410011.
%
\bibitem{R05o} S. S. Roan, The Onsager algebra symmetry of $\tau^{(j)}$-matrices in the superintegrable chiral Potts model, J. Stat. Mech. (2005) P09007; cond-mat/0505698.
%
\bibitem{R05b} S. S. Roan, Bethe ansatz and symmetry in superintegrable chiral Potts model and root-of-unity six-vertex model, in Nankai Tracts in Mathematics Vol. 10, {\it Differential Geometry and Physics}, eds. Mo-Lin Go and Weiping Zhang,  World Scientific, Singapore (2006), 399-409; cond-mat/0511543. 
%
\bibitem{R06Q} S. S. Roan, The Q-operator for root-of-unity symmetry in six vertex model, J. Phys. A: Math. Gen. 39 (2006) 12303-12325; cond-mat/0602375.
%
\bibitem{R06F} S. S. Roan, Fusion operators in the generalized $\tau^{(2)}$-model and root-of-unity symmetry of the XXZ spin chain of higher spin, J. Phys. A: Math. Theor. 40 (2007) 1481-1511; cond-mat/0607258.
%
\bibitem{R06Q8} S. S. Roan, The $Q$-operator and functional relations of the eight-vertex model at root-of-unity $\eta = 2mK/N$ for odd $N$, J. Phys. A: Math. Theor. 40 (2007) 11019-11044; cond-mat/0611316.
%
\bibitem{R07} S. S. Roan, On Q-operators of XXZ spin chain of higher spin, cond-mat/ 0702271.
%
\bibitem{R075} S. S. Roan, The transfer matrix of superintegrable chiral Potts model as the Q-operator of root-of-unity XXZ chain with cyclic representation of $U_q(sl_2)$, J. Stat. Mech. (2007) P09021; arXiv: 0705.2856.

\bibitem{TakF} L. A. Takhtadzhan and L. D. Faddeev, The quantum method of the inverse problem and Heisenberg XYZ model, Usp. Mat. Nauk 34 (1979) 13--63 (in Russian), (English Translation: Russ. Math. Surveys 34 (1979) 11--68). 
%
\bibitem{Ta} V. O. Tarasov, Cyclic monodromy matrices for the R-matrix of the six-vertex model and the chiral Potts model with fix spin boundary conditions, Intern. J. Mod. Phys. A7 Suppl. 1B (1992) 963--975.
\end{thebibliography}
\end{document}